\documentstyle[amsmath,amssymb,graphicx]{article}

\def\be{\begin{eqnarray}}
\def\ee{\end{eqnarray}}
\def\nn{\nonumber}

\def\p{\partial}
\def\tr{{\rm tr}\,}
\def\Tr{{\rm Tr}\,}

\def\Red{Reidemeister }
\def\rest{reshuffling }

\def\hor{\ \begin{array}{cc}\hline\\ \hline\end{array}\ }
\def\a{a}
\def\cob{\Bumpeq}
\def\vt{\vartheta}
\def\Q{{\mathfrak Q}}

\hoffset= - 1.0in         

\begin{document}

\hfill   ITEP/TH-37/12

\smallskip

\centerline{\Large{Introduction to Khovanov Homologies
}}
\smallskip
\centerline{\Large{I. Unreduced Jones superpolynomial}}

\bigskip

\centerline{V.Dolotin and A.Morozov}


\centerline{{\it ITEP, Moscow, Russia}}

\bigskip

\centerline{ABSTRACT}

\smallskip

\noindent
{\footnotesize
An elementary introduction to Khovanov construction of
superpolynomials.
Despite its technical complexity, this method remains
the only source of a {\it definition} of superpolynomials
from the first principles
and therefore is important for development and testing
of alternative approaches.
In this first part of the review series we concentrate
on the most transparent and unambiguous part of the story:
the unreduced Jones superpolynomials in the fundamental representation
and consider the $2$-strand braids as the main example.
Already for the $5_1$ knot the unreduced superpolynomial
contains more items than the ordinary Jones.
}


\tableofcontents

\section*{Introduction}

Knot polynomials -- the Wilson-loop averages in (refined) $3d$ Chern-Simons theory
\cite{CS,Vetco} --
are once again coming to the avant-scene of theoretical physics,
this time because they are at the intersection of a number of modern
technical developments and look as potential members of various non-trivial dualities.

Relatively well understood are the HOMFLY polynomials, basically because there
is powerful formulation in terms of the universal quantum ${\cal R}$-matrices,
which allows to pose and answer a large variety of questions,
at least in principle.
At the same time, there is a more general class of {\it superpolynomials},
for which a similarly adequate formulation is still lacking.
Instead there is a whole variety of different approaches
\cite{superpolsfirst}-\cite{superpolslast},
which give consistent results, but are still applicable
only for particular classes of knots and links.
The only exception at the moment is the original Khovanov homology method
\cite{Khof}-\cite{Khol},
which actually provides a general definition
(with relatively modest ambiguity).
Ironically, it is technically the most tricky and tedious, and its relation
to other approaches is often obscure and difficult to establish.
In addition, due to rather sophisticated presentation it is unfamiliar
to many researchers with physical background.
There was already a number of attempts to present Khovanov's approach
in simple and transparent way:  especially nice, in our opinion,
is the wonderful Bar-Natan's paper \cite{BN}.
This review note is just one more try, in fact, not very different from \cite{BN}.

In this note -- presumably the first in a series --
we discuss the simplest example of Khovanov's calculus:
for {\it unreduced} Jones superpolynomials.
"Jones" means that we are restricted to $Sl(2)$ gauge group, parameter
$A$ of HOMFLY polynomials is restricted to $A=q^2$.
Therefore we do not touch here the issue like matrix factorization,
relevant for Khovanov-Rozansky deformation for arbitrary $A=q^N$.
"Unreduced" means that $T$-deformed is the Jones polynomial itself,
not divided by the quantum dimension.
Unreduced superpolynomials are in many respects more complicated
and less interesting than the ordinary (reduced) ones,
but in Khovanov approach they are simpler, moreover,
the difference from HOMFLY, arising after $T$-deformation,
is clearly seen for them
already at the level of the simplest $2$-strand braids.

We assume some knowledge of the theory of HOMFLY polynomials,
at least of its spirit,
say, at the level of \cite{GMM}-\cite{AMMkn1}, and do not go into detail
of the corresponding motivations.
Instead we proceed directly to Khovanov's method and describe it in seven steps.
The first four of them involve reformulation of Jones polynomial as
Euler characteristic of a complex, associated with the knot/link diagram $\Gamma$,
which at the fifth step is generalized to a Poincare polynomial.
The central for these steps are the following {\it formulas}:

Step 1: Kauffman's ${\cal R}$-matrix \cite{Kauf}
\be
{\cal R}^{ij}_{kl} = q\Big(\delta^i_k\delta^j_l - q\epsilon^{ij}\epsilon_{kl}\Big), \nn \\
\Big({\cal R}^{-1}\Big)^{ij}_{kl} = -q^{-2}
\Big(\epsilon^{ij}\epsilon_{kl}-q\delta^i_k\delta^j_l\Big), \nn \\
\delta^i_i = D = q+q^{-1}
\ee

Step 2: Written in terms of this ${\cal R}$-matrix, Jones polynomial acquires a form
\be
J(q) \sim \sum_{{\rm resolutions\ of}\ \Gamma} (-q)^{|r-r_c|} D^{\nu(r)}
\ee

Step 3: On the set of resolutions one can define an action of cut-and-join operator
\be
W^\Gamma = \frac{1}{2}\sum_{a,b,c} N^\a_{bc}
\left(  p_a\frac{\p^2}{\p p_b\p p_c}
+ p_bp_c\frac{\p}{\p p_a}\right)
\label{Wop0}
\ee
where the time-variables $p_a$ are associated with the cycles, appearing in
the resolutions of $\Gamma$.
This operator acts on {\it extended} Jones polynomials and converts
them into polynomials for graphs with adjacent colorings.

Step 4: If time-variables are Miwa transformed,
auxiliary $q$-graded  vector spaces $V_a\cong V$ appear.
After that the Jones polynomial
can be rewritten as an Euler characteristic of a hypercube quiver
\be
J(q) \sim
\sum_I\, (-q)^{I}\, {\rm dim}_q C_I, \ \ \ \ \ \ \
C_I = \oplus_{\stackrel{{\rm resolutions\ of}\ \Gamma}{
{\rm with\ a\ given}\ |r-r_c|=I}}V^{\otimes \nu(r)}
\ee

Step 5: Superpolynomial is defined as Poincare polynomial of associated complex
\be
P(T,q) = \sum_{I} (qT)^{I}
\Big({\rm dim}_q {\rm Ker}(d_{I+1})
- {\rm dim}_q {\rm Im}( d_{I })\Big)
\ee

Step 6: The differentials are provided by the BRST operator,
which can be considered as a supersymmetrization of {\it a half}
of the cut-and-join operator
(\ref{Wop0}):
\be
{\Q}^\Gamma = \sum_{\stackrel{a}{b<c}}  \epsilon^a_{bc}N^\a_{bc}
\left(  Q_{ij}^k\vartheta_a^i\frac{\p^2}{\p \vartheta_b^j\p \vartheta_c^k}
+ Q^{ij}_k\vartheta_b^j\vartheta_c^k\frac{\p}{\p \vartheta_a^i}\right)
\ee
constructed with the help of the structure constants $Q$ of commutative
associative algebra.
Particular differentials are obtained by picking up some items from
${\Q}^\Gamma$, which are selected by the shape of extended Jones polynomial,
introduced at the step 3 above.

Step 7: Evaluation of the cohomologies of $d_I$.
At the present level of our understanding
this last step can be made only in particular examples,
for which we choose the simplest possible one: the $2$-strand braids,
giving the simplest possible torus links and knots, including the Hopf link
and the trefoil.
For all of them unreduced Jones polynomial are just $4$-term polynomials,
but already for the knot $5_1$ the unreduced Jones superpolynomial contains
more items (this is not the case for reduced superpolynomials: there the number
of terms becomes different and new information is provided by $T$-deformation
starting from higher number of intersections).

\section{\!\!\!step: Locality principle and Kauffman's ${\cal R}$-matrix}

Usually the classical ${\cal R}$-matrix in the fundamental representation of $Sl(2)$,
\be
{\cal R}^{ij}_{kl} = \left(\begin{array}{c|cc|c}
1 &&& \\ \hline & 0 & 1 \\ & 1 & 0 \\ \hline &&& 1
\end{array}\right) =\ \delta^i_l\delta^j_k\
=\ -\epsilon^{ij}\epsilon_{kl} + \delta^i_k\delta^j_l
\ee
is $q$-deformed into the $Sl_q(2)$ ${\cal R}$-matrix
\be
{\cal R} = \left(\begin{array}{c|cc|c} q &&& \\
\hline & q-q^{-1} & 1 &\phantom{.^{5^{5^5}}}\!\!\!\!\\
& 1 & 0 \\ \hline &&& q\end{array}\right)
\ee
which has triple eigenvalue $q$ and one $-q^{-1}$, associated with the
symmetric and antisymmetric channels $[2]$ and $[11]$ in decomposition of the
tensor product $[1]\otimes [1] = [2] + [11]$.

There is, however, another deformation \cite{Kauf}:
$$
\boxed{
{\cal R}^{ij}_{kl} = q\Big(\delta^i_k\delta^j_l - q\epsilon^{ij}\epsilon_{kl}\Big)
} \ \ \ \ \
$$
\vspace{-0.5cm}
\be
\Big({\cal R}^{-1}\Big)^{ij}_{kl} = -q^{-2}
\Big(\epsilon^{ij}\epsilon_{kl}-q\delta^i_k\delta^j_l\Big)
\label{KaRma}
\ee
where indices $i,j,k,l=1,\ldots, D$
and dimension $D$ is analytically continued
to $D=[2]_q=q+q^{-1}$.
In practice this means that
$\epsilon^{ij}\epsilon_{kj} = \delta^i_k$ and
$\epsilon^{ij}\epsilon_{ij} = \delta^i_i =D$.
(In fact this rule implies that
one can change $\epsilon^{ij}\epsilon_{kl}$ for $\delta^{ij}\delta_{kl}$
in all the expressions for link polynomials
-- this is not the same in components, but gives the same expression for traces
of ${\cal R}$-matrix products.)
The real role of $\epsilon$'s is to account for orientation change:

\begin{picture}(300,100)(-200,-50)
\put(-100,-20){\vector(1,1){40}}
\put(-60,-20){\line(-1,1){18}}
\put(-82,2){\vector(-1,1){18}}
\put(-40,0){\mbox{$=$}}
\put(-20,-20){\vector(1,1){18}}
\put(20,-20){\vector(-1,1){18}}
\put(0,0){\vector(1,1){20}}
\put(0,0){\vector(-1,1){20}}
\put(0,0){\circle*{5}}
\put(40,0){\mbox{$=$}}
%
%
%
\qbezier(140,20)(160,0)(180,20)
\qbezier(140,-20)(160,0)(180,-20)
\qbezier(70,-20)(90,0)(70,20)
\qbezier(100,20)(80,0)(100,-20)
\put(40,0){\mbox{$=$}}
\put(40,0){\mbox{$=$}}
\put(115,0){\mbox{$-q^2$}}
\put(60,0){\mbox{$q$}}
\put(158,-10){\vector(1,0){2}}
\put(162,-10){\vector(-1,0){2}}
\put(70,20){\vector(-1,1){2}}
\put(100,20){\vector(1,1){2}}
\put(140,20){\vector(-1,1){2}}
\put(180,20){\vector(1,1){2}}
\put(160,10){\circle*{3}}
\put(160,-10){\circle*{3}}
\put(158,12){\mbox{$\epsilon$}}
\put(158,-7){\mbox{$\epsilon$}}
\end{picture}
\label{epdel}

For this choice of $D$ the ${\cal R}$-matrix satisfies three
\Red conditions and skein (Hecke algebra) relation:\footnote
{
Of all these it can deserve writing the $R3$ relation in more detail.
This is the Yang-Baxter equation for braids, which is normally written as
$$
{\cal R}_{32}{\cal R}_{12}^{\pm 1} {\cal R}_{23}
= {\cal R}_{12}{\cal R}_{23}^{\pm 1} {\cal R}_{21}
$$
or, in components,
$$
\delta^i_a\left({\cal R}^{-1}\right)^{jk}_{bc}\
\left({\cal R}^{\pm 1}\right)^{ab}_{de}\delta^c_f\
\delta^d_l {\cal R}^{ef}_{mn} =
{\cal R}^{ij}_{ab}\delta^k_c\
\delta^a_d\left({\cal R}^{\pm 1}\right)^{bc}_{ef}\
\left({\cal R}^{-1}\right)^{de}_{lm}\delta^f_n
$$
Converting some indices we get:
$$
\left({\cal R}^{-1}\right)^{jk}_{bc} \left({\cal R}^{\pm 1}\right)^{ib}_{le}
{\cal R}^{ec}_{mn}
= {\cal R}^{ij}_{ab} \left({\cal R}^{\pm 1}\right)^{bk}_{en}
\left({\cal R}^{-1}\right)^{ae}_{lm}
$$
Substitution of explicit expressions (\ref{KaRma}) for ${\cal R}$-matrices gives:
$$
\Big(\delta^j_b\delta^k_c - q\delta^{jk} \delta_{bc}\Big) \delta^{ib}\delta_{le}
\Big(\delta^{ec}\delta_{mn} - q\delta^e_m\delta^c_n\Big)
= \Big(\delta^{ij}\delta_{ab} - q\delta^i_a\delta^j_b\Big)\delta^{bk}\delta_{en}
\Big(\delta^a_l \delta^e_m - q\delta^{ae}\delta_{lm}\Big),
$$
$$
\Big(\delta^j_b\delta^k_c - q\delta^{jk} \delta_{bc}\Big) \delta^i_l\delta^b_e
\Big(\delta^{ec}\delta_{mn} - q\delta^e_m\delta^c_n\Big)
= \Big(\delta^{ij}\delta_{ab} - q\delta^i_a\delta^j_b\Big)\delta^b_e\delta^k_n
\Big(\delta^a_l \delta^e_m - q\delta^{ae}\delta_{lm}\Big)
$$
and both relations should hold independently, because intermediate
${\cal R}$-matrix could enter in both powers $+1$ and $-1$.
The first of these relations is trivially satisfied.
The second one states
$$
\delta^i_l \Big((1-qD+q^2)\delta^{jk}\delta_{mn} - q\delta^j_m\delta^k_n\Big)
= \delta^k_n\Big((1-qD+q^2)\delta^{ij}\delta_{lm} - q\delta^i_l\delta^j_m\Big)
$$
what is true, provided $D = q+q^{-1} = [2]_q$.
}
\be
{\rm R_1:} & & {\cal R}^{ik}_{jk} = \delta^i_j = \Big({\cal R}^{-1}\Big)^{ik}_{jk}, \nn \\
{\rm R_2:} & & {\cal R}^{ij}_{kl}\Big({\cal R}^{-1}\Big)^{kl}_{mn} =
\delta^i_m\delta^j_n, \nn \\
{\rm R_3:} & & \left({\cal R}^{-1}\right)^{jk}_{bc} \left({\cal R}^{\pm 1}\right)^{ib}_{le}
{\cal R}^{ec}_{mn}
= {\cal R}^{ij}_{ab} \left({\cal R}^{\pm 1}\right)^{bk}_{en}
\left({\cal R}^{-1}\right)^{ae}_{lm}, \nn \\
{\rm skein:} & & q^{-2}{\cal R}^{ij}_{kl} - q^{2}\Big({\cal R}^{-1}\Big)^{ij}_{kl} =
-(q-q^{-1})\delta^i_k\delta^j_l
\ee

This means that such ${\cal R}$ can be used to evaluate the
unreduced Jones polynomial for arbitrary like diagram, i.e. the
Wilson-loop average in $3d$ Chern-Simons theory with the gauge group $Sl(2)$
\cite{CS} in the temporal gauge $A_0=0$ \cite{MSm}.
Projection of a
knot/link on a $2d$ plane provides a link diagram,
which is a planar $4$-valent graph with two types of vertices,
which we denote as black and white.
One puts ${\cal R}$ at black vertices and ${\cal R}^{-1}$ at the white ones
and contracts indices along all edges.
This provides a knot polynomial (Jones in the fundamental representation
for above choice of ${\cal R}$), which is independent of the projection --
a topological invariant.
Invariance is a corollary of \Red relations, and the possibility to apply
them is that above construction -- contraction of ${\cal R}$ matrices --
is {\it local}: one can apply an identity, valid for several concrete vertices,
and it will continue to hold for entire graph.
This locality property looks absolutely trivial in this formulation,
but it turns into not-quite-a-trivial-theorem in reformulations below.
What is important, {\it locality} is the property of the {\it construction},
it does not depend on the nature of particular invariance:
one could use locality with whatever one likes, not obligatory \Red moves.
On the contrary, \Red invariance
is a property of particular ${\cal R}$-matrix,
and it is a {\it local} condition, imposed at the level of one, two, and three
vertices -- after that {\it locality} allows to extend it to arbitrary graphs.

For example, for $2$-strand braid with $n$ crossings
(it is a knot and a two-component link for $n$ odd and even respectively)
we have (the picture is rotated by $90^\circ$ to save the space, the same is done
in the first part of the formula):

\begin{picture}(300,110)(-150,-55)
\qbezier(0,0)(-10,10)(-20,10)
\qbezier(0,0)(-10,-10)(-20,-10)
\qbezier(0,0)(20,20)(40,0)
\qbezier(40,0)(60,20)(80,0)
\qbezier(0,0)(20,-20)(40,0)
\qbezier(40,0)(60,-20)(80,0)
\qbezier(80,0)(90,10)(100,10)
\qbezier(120,10)(130,10)(140,0)
\qbezier(140,0)(160,20)(180,0)
\qbezier(180,0)(190,10)(200,10)
\qbezier(80,0)(90,-10)(100,-10)
\qbezier(120,-10)(130,-10)(140,0)
\qbezier(140,0)(160,-20)(180,0)
\qbezier(180,0)(190,-10)(200,-10)
\put(105,0){\mbox{$\ldots$}}
\put(0,0){\circle*{5}}
\put(40,0){\circle*{5}}
\put(80,0){\circle*{5}}
\put(140,0){\circle*{5}}
\put(180,0){\circle*{5}}
%
\put(20,10){\vector(1,0){2}}
\put(60,10){\vector(1,0){2}}
\put(100,10){\vector(1,0){2}}
\put(120,10){\vector(1,0){2}}
\put(160,10){\vector(1,0){2}}
\put(20,-10){\vector(1,0){2}}
\put(60,-10){\vector(1,0){2}}
\put(100,-10){\vector(1,0){2}}
\put(120,-10){\vector(1,0){2}}
\put(160,-10){\vector(1,0){2}}
%
\qbezier(-20,10)(-80,20)(-20,40)
\qbezier(200,10)(260,20)(200,40)
\qbezier(-20,40)(90,70)(200,40)
\qbezier(-20,-10)(-80,-20)(-20,-40)
\qbezier(200,-10)(260,-20)(200,-40)
\qbezier(-20,-40)(90,-70)(200,-40)
\put(90,55){\vector(-1,0){2}}
\put(90,-55){\vector(-1,0){2}}
\end{picture}

$$
J^{(n)}_{_\Box}(q) = \Tr{\cal R}^n = \underbrace{
{\cal R}^{ij}_{kl}{\cal R}^{kl}_{mn}\ldots {\cal R}^{pq}_{ij}}_n =
q^n\Tr\!\!
\left\{\left(
\ \Big|\ \Big|\, - q \,\begin{array}{c}\hline\\ \hline\end{array}\
\right)^n\right\}
= q^n\left\{ \begin{array}{c}O\\O\end{array}\! - nq\cdot O + \frac{n(n-1)}{2}\,q^2\cdot O^2
+ \ldots \right\} =
$$
\vspace{-0.5cm}
\be
=q^n \Big(D^2-1 + (1-qD)^n\Big) = q^n\Big(q^{-2}+1+q^{2} + (-q^2)^n\Big)
\label{Jn}
\ee
where the circle diagram stands for the trace of unity, $\Tr I = D = q+q^{-1}$.
Similarly,
\be
\Tr\Big({\cal R}^{-1}\Big)^n = \left(-\frac{1}{q^2}\right)^n \Big\{(D-q)^n +
(-q)^n(D^2-1)\Big\}
= \frac{1}{q^n}\left\{q^2+1+q^{-2} + \left(-\frac{1}{q^2}\right)^n\right\}
= J^{(n)}_{_\Box}(q^{-1})
\ee
The two expressions are related by $Z_2$-symmetry $q\longrightarrow q^{-1}$.
One can notice that for any $n$ the polynomials at the r.h.s. are divisible by
$D = J^{\rm unknot}_{_\Box}$
(though in a different ways for knots and links), what allows to introduce the
{\it reduced} Jones polynomial $\check J_R = J_R/J_R^{\rm unknot}$,
which plays a big role in the theory of knot polynomials,
but will be not discussed in the present paper.
As for all torus knots, the {\it unreduced} Jones polynomial consists of just
four items
(this property gets obvious in alternative -- matrix model -- representation, see
the last paper in \cite{RJ} and the very first in \cite{AMMkn1}).

\section{\!\!\!step: The double polynomial and the cube of resolutions
\label{ste2}}

Concrete form of the Kauffman ${\cal R}$-matrix (\ref{KaRma}) implies that the
Jones polynomial is actually a polynomial in two (related) variables $q$ and $D$,
which possesses a nice pictorial representation, which we actually used
in the above example.
The convoluted product of ${\cal R}$-matrices for a planar oriented $4$-valent graph
$\Gamma_c$ with black and white vertices\footnote{
In knot-theory applications the graphs have all these properties,
but many elements of the construction are more general: the graphs need
not be planar, and also can have more types of vertices of different valencies,
though in this case {\it resolutions} will be substituted by other notions.
}
is represented as a sum over all possible {\it resolutions}
$r$ of the graph:
\be
\boxed{
J^{\Gamma_c}_{_\Box}(q) = (-)^{n_\circ}q^{n_\bullet-2n_\circ}
\sum_r^{2^n} (-q)^{|r-r_c|} D^{\nu(r)}
}
\label{qDpol}
\ee
Here $n = n_\bullet+n_\circ$ is the number of vertices in $\Gamma_c$,
and each of the $2^n$ resolutions
contributes a product of two factors:
$D^{\nu(r)}$, where $\nu(r)$ is the number of connected components in the
resolved graph, and $(-q)^{|r-r_c|}$ where $|r-r_c|$ is the number of flips,
needed to achieve the given resolution $r$ from the original $r_c$,
obtained when all black vertices are represented by $\Big|\ \Big|$
and all white vertices -- by $\hor$.
This distance is the only place in this approach,
where the colors of vertices play a role: changing colors imply that
we count the number of flips $|r-r_c|$ from another $r_c$.

Thus what matters in (\ref{qDpol}) is the net of resolutions, connected by
elementary flips $\ \Big|\ \Big| \ \leftrightarrow\ \hor$.
Clearly, this net is nothing but an $n$-dimensional hypercube
with $2^n$ vertices and $2^{n-1}n$ edges.
Vertices are labeled by $n$-digit binary numbers
$\alpha =[\alpha_1\alpha_2\ldots \alpha_n]$, with
$\alpha_1,\ldots,\alpha_n=0,1$.
If we forget the colors of vertices, $\Gamma_c\longrightarrow \Gamma$
and enumerate the vertices of the graph $\Gamma$, then
over the vertex $\alpha$ we put a particular resolution of the graph $\Gamma$,
where at $I$-th vertex ($I=1,\ldots,n$)
we put $\ \Big|\ \Big|\ $ if $\alpha_I=0$ and put $\hor $ if $\alpha_I=1$.
The $n$ edges, connecting a given vertex of the hypercube to adjacent ones,
describe elementary flips at arbitrary vertex $J$ of $\Gamma$.
Original graph $\Gamma_c$ with colored vertices defines a distinguished
vertex in the hypercube with a resolution $r_c$,
then $|r-r_c| =  \sum_i \big|\,\alpha_I-\alpha^c_I\,\big|$, and now all the edges
acquire orientation: they point away from $r_c$,
and this is a step towards defining
a quiver structure on the hypercube
(a very simple one, in fact -- just brushing from one vertex of the opposite).

\bigskip

{\bf Example, $n=1$:}

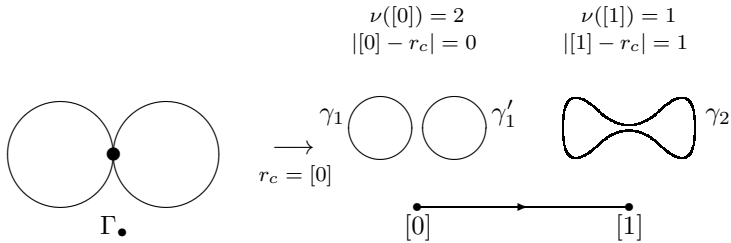
\begin{picture}(300,150)(-150,-90)
\put(-120,0){\circle{40}}
\put(-80,0){\circle{40}}
\put(-100,0){\circle*{5}}
\put(-40,0){\mbox{$\longrightarrow$}}
\put(15,-20){\line(1,0){80}}
\put(15,-20){\circle*{3}}
\put(95,-20){\circle*{3}}
\put(55,-20){\vector(1,0){2}}
\put(1,10){\circle{25}}
\put(29,10){\circle{25}}
 \qbezier(70,10)(70,30)(85,15)
\qbezier(70,10)(70,-10)(85,5)
\qbezier(85,15)(95,7)(105,15)
\qbezier(85,5)(95,13)(105,5)
\qbezier(120,10)(120,30)(105,15)
\qbezier(120,10)(120,-10)(105,5)
\put(10,-30){\mbox{[0]}}
\put(90,-30){\mbox{[1]}}
\put(-105,-30){\mbox{$\Gamma_\bullet$}}
\put(-3,50){\mbox{{\footnotesize $\nu([0])=2$}}}
\put(77,50){\mbox{{\footnotesize $\nu([1])=1$}}}
\put(-10,40){\mbox{{\footnotesize $|[0]-r_c|=0$}}}
\put(70,40){\mbox{{\footnotesize $|[1]-r_c|=1$}}}
\put(-45,-10){\mbox{{\footnotesize $r_c = [0]$}}}
\put(-100,-60){\mbox{$J^\bullet_{_\Box}(q) = q(D^2-qD) = D, \ \ \ \ \ \
\check J^\bullet_{_\Box}(q) = \frac{J^\bullet_{_\Box}(q)}{J^{\rm unknot}_{_\Box}(q)} = 1$}}
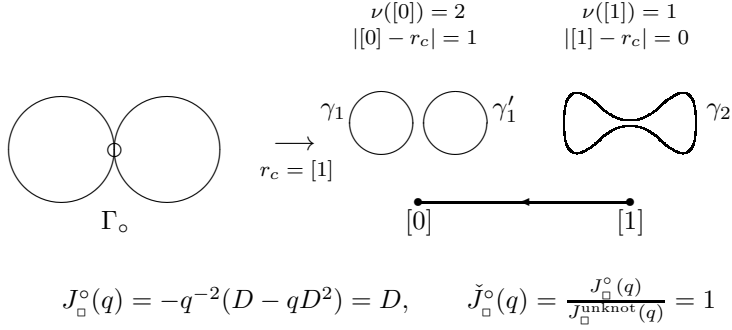
\put(-22,13){\mbox{$\gamma_1$}}
\put(43,13){\mbox{$\gamma_1'$}}
\put(124,13){\mbox{$\gamma_2$}}
\end{picture}

\vspace{2cm}

\begin{picture}(300,110)(-150,-45)
\put(-120,0){\circle{40}}
\put(-80,0){\circle{40}}
\put(-100,0){\circle{5}}
\put(-40,0){\mbox{$\longrightarrow$}}
\put(15,-20){\line(1,0){80}}
\put(15,-20){\circle*{3}}
\put(95,-20){\circle*{3}}
\put(55,-20){\vector(-1,0){2}}
\put(1,10){\circle{25}}
\put(29,10){\circle{25}}
\qbezier(70,10)(70,30)(85,15)
\qbezier(70,10)(70,-10)(85,5)
\qbezier(85,15)(95,7)(105,15)
\qbezier(85,5)(95,13)(105,5)
\qbezier(120,10)(120,30)(105,15)
\qbezier(120,10)(120,-10)(105,5)
\put(10,-30){\mbox{[0]}}
\put(90,-30){\mbox{[1]}}
\put(-105,-30){\mbox{$\Gamma_\circ$}}
\put(-3,50){\mbox{{\footnotesize $\nu([0])=2$}}}
\put(77,50){\mbox{{\footnotesize $\nu([1])=1$}}}
\put(-10,40){\mbox{{\footnotesize $|[0]-r_c|=1$}}}
\put(70,40){\mbox{{\footnotesize $|[1]-r_c|=0$}}}
\put(-45,-10){\mbox{{\footnotesize $r_c = [1]$}}}
\put(-120,-60){\mbox{$J^\circ_{_\Box}(q) = -q^{-2}(D-qD^2) = D, \ \ \ \ \ \
\check J^\circ_{_\Box}(q) = \frac{J^\circ_{_\Box}(q)}{J^{\rm unknot}_{_\Box}(q)} = 1$}}
\put(-22,13){\mbox{$\gamma_1$}}
\put(43,13){\mbox{$\gamma_1'$}}
\put(124,13){\mbox{$\gamma_2$}}
\end{picture}

\vspace{2cm}

{\bf Example, $n=2$:}

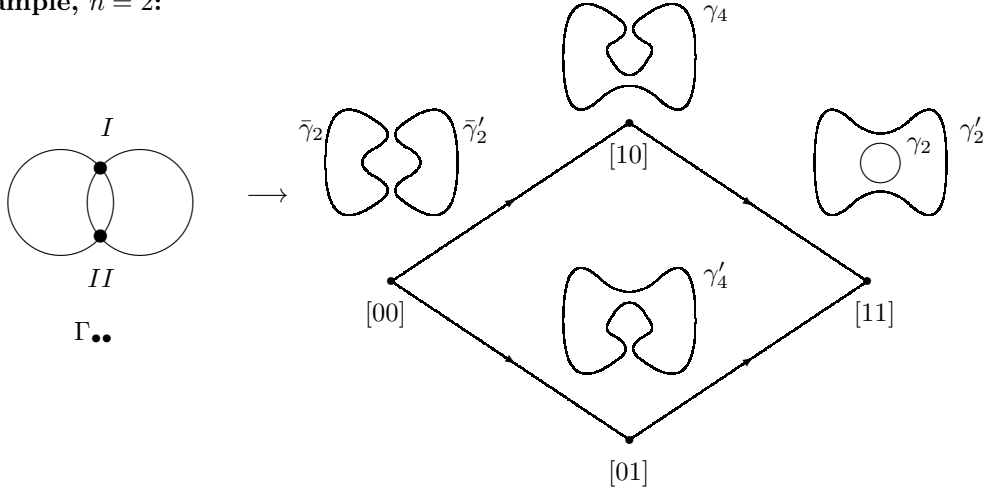
\begin{picture}(300,170)(-150,-100)
\put(-110,0){\circle{40}}
\put(-80,0){\circle{40}}
\put(-95,13){\circle*{5}}
\put(-95,-13){\circle*{5}}
\put(-95,25){\mbox{$I$}}
\put(-100,-32){\mbox{$II$}}
\put(-105,-52){\mbox{$\Gamma_{\bullet\bullet}$}}
\put(-40,0){\mbox{$\longrightarrow$}}
\qbezier(15,-30)(60,0)(105,30)
\qbezier(15,-30)(60,-60)(105,-90)
\qbezier(195,-30)(150,0)(105,30)
\qbezier(195,-30)(150,-60)(105,-90)
\put(15,-30){\circle*{3}}
\put(195,-30){\circle*{3}}
\put(105,30){\circle*{3}}
\put(105,-90){\circle*{3}}
\put(60,0){\vector(3,2){2}}
\put(150,0){\vector(3,-2){2}}
\put(60,-60){\vector(3,-2){2}}
\put(150,-60){\vector(3,2){2}}
\put(5,-45){\mbox{[00]}}
\put(190,-45){\mbox{[11]}}
\put(97,15){\mbox{[10]}}
\put(97,-105){\mbox{[01]}}
\put(200,15){\circle{15}}
\qbezier(175,15)(175,45)(190,30)
\qbezier(175,15)(175,-15)(190,0)
\qbezier(190,30)(200,22)(210,30)
\qbezier(190,0)(200,8)(210,0)
\qbezier(225,15)(225,45)(210,30)
\qbezier(225,15)(225,-15)(210,0)
%
%
%
%
\qbezier(80,55)(80,85)(100,70)
\qbezier(80,55)(80,25)(95,40)
\qbezier(95,40)(105,48)(115,40)
\qbezier(130,55)(130,85)(110,70)
\qbezier(130,55)(130,25)(115,40)
\qbezier(98,54)(105,42)(112,54)
\qbezier(100,70)(108,65)(98,60)
\qbezier(98,54)(95,58)(98,60)
\qbezier(110,70)(102,65)(112,60)
\qbezier(112,54)(115,58)(112,60)
\qbezier(80,-45)(80,-75)(100,-60)
\qbezier(80,-45)(80,-15)(95,-30)
\qbezier(95,-30)(105,-38)(115,-30)
\qbezier(130,-45)(130,-75)(110,-60)
\qbezier(130,-45)(130,-15)(115,-30)
\qbezier(98,-44)(105,-32)(112,-44)
\qbezier(100,-60)(108,-55)(98,-50)
\qbezier(98,-44)(95,-48)(98,-50)
\qbezier(110,-60)(102,-55)(112,-50)
\qbezier(112,-44)(115,-48)(112,-50)
\qbezier(-10,15)(-10,45)(10,30)
\qbezier(-10,15)(-10,-15)(10,0)
\qbezier(40,15)(40,45)(20,30)
\qbezier(40,15)(40,-15)(20,0)
\qbezier(10,30)(18,25)(8,20)
\qbezier(10,0)(18,5)(8,10)
\qbezier(20,30)(12,25)(22,20)
\qbezier(20,0)(12,5)(22,10)
\qbezier(8,10)(0,15)(8,20)
\qbezier(22,10)(30,15)(22,20)
\qbezier(100,-60)(108,-55)(98,-50)
\qbezier(100,-60)(108,-55)(98,-50)
\qbezier(100,-60)(108,-55)(98,-50)
\qbezier(100,-60)(108,-55)(98,-50)
\put(-20,25){\mbox{$\bar\gamma_2$}}
\put(42,25){\mbox{$\bar\gamma_2'$}}
\put(133,70){\mbox{$\gamma_4$}}
\put(133,-30){\mbox{$\gamma_4'$}}
\put(210,20){\mbox{$\gamma_2$}}
\put(230,25){\mbox{$\gamma_2'$}}
\end{picture}

\bigskip

In Appendix A at the end of this paper we explain, how the topological
invariance of the answer (\ref{qDpol}) for Jones polynomial can be proved in
terms of this hypercube construction -- without reference to
the algebraic proof in sec.1.
Since algebraic proof exists and is far more straightforward,
there is no direct need to consider more sophisticated alternatives --
but they help to highlight other sides of the theory,
thus we include this alternative proof, but aside of the main
line of our presentation.
As a clarifying example,
in Appendix A we discuss briefly what happens if one attempts to  include
the third ("$u$-channel") resolution: the result remains invariant under the two
of the three \Red shifts $E1$ and $R2$, but $R3$ -invariance is violated
(at the ${\cal R}$-matrix language of sec.1 this simply means that
the Yang-Baxter relation is violated).

\bigskip

Given (\ref{qDpol}), it is a natural idea to release $q$-dependence of $D$
and treat $q$ and $D$ as independent parameters.
Unfortunately, it does not work so simple: if we do this directly in (\ref{qDpol})
topological invariance under \Red moves will be lost: say,
\be
\pi(t,D) = \sum_r^{2^n} t^{|r-r_c|} D^{\nu(r)}
\ee
is not an invariant.
However, after profound modification of (\ref{qDpol}) this idea turns viable
and provides an invariant two-parametric generalization of Jones polynomial.
Still three more steps need to be made for this.
The first of them will extend Jones polynomial even further
(sacrificing topological invariance), but
reveal a new hidden structure
of auxiliary vector spaces at the hypercube vertices.
Then, after a kind of a supersymmetrization, we return back to
topological invariant quantities -- with a new deformation parameter gained
in the process.

\section{\!\!\!step: Cobordisms, cut-and-join operator and {\it extended}
Jones
}

Resolution $r$ of the graph $\Gamma$ substitutes it by a set of
disconnected cycles $\gamma_a$, $a\in r$,
where $a$ is some label, used to enumerate all cycles, arising in
all possible resolutions.
Each cycle had a length $n_a$ (consists of $n_a$ edges),
and there is an obvious constraint
\be
\sum_{a\in r} n_a = n \ \ \ \ \ \ \ \
\forall\ \ {\rm resolution}\ r \ {\rm of}\ \Gamma
\ee
Also $\gamma_a \cap \gamma_b = \emptyset$, i.e. the two cycles
do not contain common edges, if
$\gamma_a$ and $\gamma_b$ belong to the same resolution $r$, $a,b\in r$.
At the same time, particular $\gamma_a$ can participate in
different resolutions $r$ of $\Gamma$.

Elementary flips $\ \Big|\ \Big| \ \longleftrightarrow \ \hor\ $
act on the set of cycles $\{\gamma_a\}$, either gluing two into one
or splitting one into two:
\be
\gamma_a = \gamma_b \cob \gamma_c,
\ \ \ \ \ \ \ \ \ \ \ \ \ \ n_a = n_b+n_c
\ee
and we introduce the cobordism structure constant $N^\a_{bc}$,
which is equal to one, if a flip exists, connecting the triple $(a|bc)$,
and zero otherwise. By this definition, $N^a_{bc} = N^a_{cb}$, but
$N^a_{bc}=0$ for $b=c$.
There are "undecomposable" cycles $\gamma_a$, for which all
$N^a_{bc} =0$, and a sequence of flips can be used to decompose
arbitrary flip into such "minimal" ones -- but we do not use
this additional structure in the present paper.

In order to develop a more algebraic formalism, we associate with
each cycle $\gamma_a$ a "time-variable" $p_a$.
Conceptually, it can be interpreted as a Wilson-loop variable
$p_a = \tr U^{n_a}$ with some matrix abstract matrix $U$,
defined on the edges of $\Gamma$ (in the simplest case, just constant)
-- but again we do not follow this line in the present text.
Instead we proceed to the hypercube.
With each  hypercube vertex, i.e. with every resolution $r$
we can now associate a monomial $p\,^r \equiv \prod_{a\in r} p_a$
as a product of all the $\nu(r)$ time-variables, associated with cycles which
participate in the resolution $r$.
Edges of the hypercube correspond to the elementary flips
between "adjacent" resolutions,\footnote{
The fact that there are exactly $n$ edges, terminating in each vertex
and, more generally, that edges form a hypercube, are strong restrictions
on the set of the structure constants $\{N^\a_{bc}\}$,
which are automatically satisfied, if we build operator $W$, starting from
some graph $\Gamma$.}
and each such flip is associated with exactly one item in the
{\it cut-and-join operator}
\be
\boxed{
W^\Gamma = \frac{1}{2}\sum_{a,b,c} N^\a_{bc}
\left(  p_a\frac{\p^2}{\p p_b\p p_c}
+ p_bp_c\frac{\p}{\p p_a}\right)
}
\label{Wop}
\ee
(note that there are no factors like $n_bn_c$ or $n_a=n_b+n_c$ in this formula,
as in more familiar cut-and-join operators of \cite{MMN} -- instead
there can be numerous $p_a$ with the same $a$).

As before, the coloring of the vertices in the graph $\Gamma_c$ defines the "starting"
vertex $r_c$ of the hypercube and al other vertices acquire the numbers
$|r-r_c|$, equal to their distance from $r_c$
(i.e. the number of edges, connecting them to $v_c$).
Also all edges of the hypercube turn into arrows, pointing away from $r_c$
-- and for each triple $(a|bc)$, contributing to (\ref{Wop}) one of the two terms
is selected from the cut-and-join operator.

The last thing to do at this step is to introduce the
"extended Jones polynomial"
(compare with extended HOMFLY of \cite{AMMkn1}):
\be
\boxed{
{\cal J}^{\Gamma_c}\{t|p_k^\a\} =
\sum_{r \in {\rm cypercube}} t^{|r-r_c|} \prod_{a\in r}^{\nu(r)} p_a
}
\ee
It lives on the infinite-dimensional space of time-variables $p_a$,
and can be a subject of character calculus and
can be analyzed from the perspectives of integrability
($\tau$-functions) theory.
It, however, is not a real topological invariant:
the topological invariant Jones polynomial arises after restriction
to {\it topological locus}
\be
p_a = D = q+q^{-1}
\label{tolo}
\ee
and for $t=-q$:
\be
J^{\Gamma_c}(q) = q^{n_\bullet}(-1/q^2)^{n_\circ}
{\cal J}^{\Gamma_c}\Big\{t=-q\Big|p_a = D\Big\}
\label{JcalJ}
\ee

\bigskip

Before proceeding further, we consider examples
-- the same as in s.\ref{ste2}.

\bigskip

{\bf Example: $n=1$}

In this case we have just a single cycle of length $2$ and two cycles of length $1$.
This time the hypercube is a segment with two vertices and
$p^{[0]} = p_1p_1'$, $p^{[1]} = p_2$.
Depending on the color of the vertex of original graph we get two
extended Jones polynomials
\be
{\cal J}^\bullet = p_1p_1' + tp_2
\ee
and
\be
{\cal J}^\circ = p_2+tp_1p_1'
\ee
interchanged by the action of cut-and-join operator
\be
\hat W = p_2\frac{\p^2}{\p p_1\p p_1'} +  p_1^2\frac{\p}{\p p_2}
\ee
\be
\hat W {\cal J}^\bullet = {\cal J}^\circ, \ \ \ \
\hat W {\cal J}^\circ = {\cal J}^\bullet
\ee
The ordinary unreduced Jones polynomials appear according to the rule (\ref{JcalJ}):
\be
J^\bullet(q) = q(D^2-qD) = D, \nn \\
J^\circ(q) = (-1/q^2)(D-qD^2) = D
\ee

\bigskip

{\bf Example: $n=2$}

This time we have two variable $p_4, p_4'$ and four variables $\bar p_2,\bar p_2',p_2,p_2'$
with decomposition rules encoded in the operator
\be
\hat W =   (p_4+p_4')\left(\frac{\p^2}{\p \bar p_2\p \bar p_2'}
+  \frac{\p^2}{\p p_2\p p_2'}\right)
+  (\bar p_2\bar p_2'+p_2p_2')\left(\frac{\p}{\p p_4} + \frac{\p}{\p p_4'}\right)
\ee
The four extended Jones polynomials
\be
{\cal J}^{\bullet\bullet} = \bar p_2\bar p_2' + t(p_4+p_4') + t^2p_2p_2', \nn \\
{\cal J}^{\bullet\circ} = p_4 + t(\bar p_2\bar p_2'+p_2p_2') + t^2p_4', \nn\\
{\cal J}^{\circ\bullet} = p_4' + t(\bar p_2\bar p_2'+p_2p_2') + t^2p_4, \nn \\
{\cal J}^{\circ\circ} = p_2p_2' + t(p_4+p_4') + t^2\bar p_2\bar p_2'
\ee
are again intertwined by $\hat W$:
\be
\hat W {\cal J}^{\bullet\bullet} = {\cal J}^{\bullet\circ} + {\cal J}^{\circ\bullet}
= \hat W {\cal J}^{\circ\circ},\nn \\
\hat W {\cal J}^{\bullet\circ} = {\cal J}^{\bullet\bullet} + {\cal J}^{\circ\circ}
= \hat W {\cal J}^{\circ\bullet}
\ee

\bigskip

{\bf Example: $n=3$, trefoil}

The set of cycles consists of three $p_6,p_6',p_6''$,
three $p_4,p_4',p_4''$, two $p_3,p_3'$
and three $p_2,p_2',p_2''$,
related through
\be
\hat W =
p_6\left(\frac{\p^2}{\p p_4'\p p_2'} + \frac{\p^2}{\p p_4''\p p_2''}\right)
+ p_6'\left(\frac{\p^2}{\p p_4\p p_2} + \frac{\p^2}{\p p_4''\p p_2''}\right)
+ p_6''\left(\frac{\p^2}{\p p_4\p p_2} + \frac{\p^2}{\p p_4'\p p_2'}\right)
+ \nn \\
+  (p_6+p_6'+p_6'')\frac{\p^2 }{\p p_3\p p_3'}
+   \left(p_4\frac{\p^2}{\p p_2'\p p_2''} + p_4'\frac{\p^2}{\p p_2\p p_2''} +
p_4''\frac{\p^2}{\p p_2\p p_2'}\right) + \nn \\
+  \left( (p_4'p_2' + p_4''p_2'')\frac{\p}{\p p_6} +
(p_4p_2 + p_4''p_2'')\frac{\p}{\p p_6'}
+ (p_4p_2 + p_4'p_2')\frac{\p}{\p p_6''}\right)+ \nn \\
+  p_3p_3'\left(\frac{\p}{\p p_6} + \frac{\p}{\p p_6'}
 + \frac{\p}{\p p_6''}\right)
+  \left(p_2'p_2''\frac{\p}{\p p_4} + p_2p_2''\frac{\p}{\p p_4'} +
p_2p_2'\frac{\p}{\p p_4''}\right)
\ee
The corresponding extended Jones polynomials are:
\be
{\cal J}^{\bullet\bullet\bullet} = p_3p_3' + t(p_6+p_6'+p_6'') +
t^2(p_4p_2+p_4'p_2'+p_4''p_2'') + t^3p_2p_2'p_2'', \nn \\
{\cal J}^{\circ\bullet\bullet} = p_6 + t( p_4'p_2' + p_4''p_2'' + p_3p_3') +
t^2(p_6'+p_6''+p_2p_2'p_2'') + t^3p_4p_2, \nn \\
{\cal J}^{\bullet\circ\bullet} = p_6' + t( p_4p_2 + p_4''p_2'' + p_3p_3') +
t^2(p_6+p_6''+p_2p_2'p_2'') + t^3p_4'p_2', \nn \\
{\cal J}^{\bullet\bullet\circ} = p_6'' + t( p_4p_2 + p_4'p_2' + p_3p_3') +
t^2(p_6+p_6'+p_2p_2'p_2'') + t^3p_4''p_2'', \nn \\
{\cal J}^{\bullet\circ\circ} = p_4p_2 + t(p_6'+p_6''+p_2p_2'p_2'')
+t^2 (p_4'p_2' + p_4''p_2'' + p_3p_3') + t^3p_6, \nn \\
{\cal J}^{\circ\bullet\circ} = p_4'p_2' + t(p_6+p_6''+p_2p_2'p_2'')
+t^2 (p_4p_2 + p_4''p_2'' + p_3p_3') + t^3p_6', \nn \\
{\cal J}^{\circ\circ\bullet} = p_4''p_2'' + t(p_6+p_6'+p_2p_2'p_2'')
+t^2 (p_4p_2 + p_4'p_2' + p_3p_3') + t^3p_6'', \nn \\
{\cal J}^{\circ\circ\circ} = p_2p_2'p_2'' + t(p_4p_2+p_4'p_2'+p_4''p_2'') +
t^2(p_6+p_6'+p_6'') + t^3p_3p_3'
\label{Jontre}
\ee
and cut-and-join operator acts between them as follows:
\be
\hat W {\cal J}^{\bullet\bullet\bullet} =
{\cal J}^{\circ\bullet\bullet}
+ {\cal J}^{\bullet\circ\bullet} + {\cal J}^{\bullet\bullet\circ},\nn \\
\hat W {\cal J}^{\circ\bullet\bullet} =
{\cal J}^{\bullet\bullet\bullet}
+ {\cal J}^{\circ\circ\bullet} + {\cal J}^{\circ\bullet\circ}, \nn \\
\!\!\!\!\!\!\!\!\!\!\!\!\!\!\ldots
\ee

Clearly, the action of $\hat W$ on ${\cal J}^{\Gamma_c}$ changes the colors of
vertices in the graph:
\be
\hat W {\cal J}^{\Gamma_c} = \sum_{c': \ |r_{c'}-r_c|=1} {\cal J}^{\Gamma_{c'}}
\ee
and the sum is over all starting vertices in the hypercube,
adjacent to $r_c$.

\section{\!\!\!step. Towards quiver: from resolutions to vector spaces}

The rule (\ref{tolo}) is in fact very suggestive:
usually \cite{DMMSS,AMMkn1} topological locus is associated with
putting $X=I$ in the Miwa transform $p_a = \tr X^{n_a}$.
Perhaps, surprisingly, (\ref{tolo}) does not depend on $n_a$ --
this remains to be understood,  but somehow $n_a$ does not seem to play any role
in the topological cobordism construction, underlying Khovanov's approach.
Anyhow, what is beyond doubt, is the re-appearance of the $D$-dimensional
(or $q$-graded $2$-dimensional, if one prefers) vector space:
it was actually present at the very beginning, at step 1,
disappeared at steps 2 and 3 and is now back through the Miwa transform.

\bigskip

To be precise, instead of $\prod_{a\in r} p_a$ at every vertex
of the hypercube we now put a tensor product of vector spaces
$\otimes_{a\in r}^{\nu(r)} V_a$, all $V_a$ isomorphic to the $q$-graded
two-dimensional space $V$,  $\ V_a \cong V$.
Now we can rewrite (\ref{qDpol}) as
\be
\boxed{
J^{\Gamma_c}_{_\Box}(q) = (-)^{n_\circ}q^{n_\bullet-2n_\circ}
\sum_r^{2^n} (-q)^{|r-r_c|}\ {\rm dim}_q\! \left(V^{\otimes\nu(r)}\right)
= (-)^{n_\circ}q^{n_\bullet-2n_\circ}\sum_{I=0}^n\, (-q)^I\, {\rm dim}_q C_I
}
\label{qDpoldim}
\ee
where the new vector spaces $C_I$ are direct sums of all the vector
spaces, associated with all the hypercube vertices, lying at a given
distance from initial vertex $r_0$:
\be
C_I = \oplus_{r:\ |r-r_c|=I}\ V^{\otimes\nu(r)}
\ee
To define quantum dimension it is enough to fix a basis in $V$ so that
the two basis vectors $v_\pm$ have weights $q^{\pm 1}$.

This ends Kauffman's reformulation of the ordinary Jones polynomial,
and brings it to the form, allowing Khovanov's deformation.
In original formulation topological invariance followed immediately from
the obvious locality of the contracted product of ${\cal R}$-matrices
and from the properties of Kauffman's ${\cal R}$-matrices
(behavior under three \Red moves).
In (\ref{qDpoldim}) locality is less transparent.
The idea now is that topological information is now encoded
in \rest properties of the graph under the elementary flips
of resolutions at particular vertices.
While flips are local, {\rest}s are not.
By adding a new vertex, like in the \Red move $R_1$
we add a new dimension to the hypercube,
and locality should be now treated in homological terms.
But for this to work one needs to reformulate
(\ref{qDpoldim}) in terms of cohomologies.

For this we should recall that there are operators acting along the
hypercube edges and lift them to operators, acting between tensor products
of vector spaces $V$.
After that -- if we are lucky -- the linear combinations of these operators,
arising in projection to the spaces $C_I$, will provide {\it differentials}
(possess the property $d_{i+1}d_I = 0$), allowing a cohomological
reformulation of (\ref{qDpoldim}).

It turns out that the differentials are indeed produced by a kind of
supersymmetrization of the cut-and-join operator (\ref{Wop}),
but before we switch to this issue (step 6), we describe what we need
it for.

\section{\!\!\!step. Restricting to kernels}

Eq.(\ref{qDpoldim}) already looks like an Euler characteristic
of the graded complex
\be
{\cal C}:\ \ \ \
0 \longrightarrow C_0 \stackrel{d_1}{\longrightarrow} C_1 \stackrel{d_2}{\longrightarrow}
\ldots \stackrel{d_{n}}{\longrightarrow} C_n \longrightarrow 0
\ee
and Euler characteristic can be alternatively rewritten in terms of cohomologies:
\be
J^{\Gamma_c}_{_\Box}(q) =
(-)^{n_\circ}q^{n_\bullet-2n_\circ}\sum_{I=0}^n\, (-q)^I\, {\rm dim}_q C_I
= (-)^{n_\circ}q^{n_\bullet-2n_\circ}\sum_{I=0}^n\, (-q)^I\, {\rm dim}_q H_I =\nn \\
= \sum_{I=0}^n (-q)^I \left\{{\rm dim}_q \Big({\rm Ker}(d_{I+1})\Big)
- {\rm dim}_q\Big({\rm Im}(d_{I})\Big)\right\}
\label{Jopo}
\ee
where at the last stage all the $q$-factors are absorbed into regrading of the
$C_I$ spaces.

This formula already allows a $T$-deformation:
the Jones superpolynomial
\be
\boxed{
P^{\Gamma_c}_{_\Box}(T,q) = q^{n_\bullet}\cdot(Tq^2)^{-n_\circ}
\sum_{I=0}^n (qT)^I \left\{{\rm dim}_q \Big({\rm Ker}(d_{I+1})\Big)
- {\rm dim}_q\Big({\rm Im}(d_{I})\Big)\right\}
}
\label{poinc}
\ee
is invariant under \Red moves, provided the differentials $d_J$ are chosen in
appropriate way.

Again, invariance is a two-level statement:

(i) the Poincare polynomial (\ref{poinc}) is {\it local} in the sense that
it is not changed if the complex ${\cal C}$ is substituted by ${\cal C}'$
such that either the factor-complex ${\cal C}/{\cal C}'$ is acyclic
(has vanishing cohomologies) or ${\cal C'} = {\cal C}/{\cal C}''$ with
acyclic ${\cal C}''$,

(ii) differentials $d_I$ are such, that the changes of complex ${\cal C}$,
associated with the \Red moves, provide acyclic factor-complexes.

Reverting colors of all vertices, $\Gamma_c \longrightarrow \Gamma_{\bar c}$,
reverses orientation of the link,
and the superpolynomial changes in a simple way:
\be
P^{\Gamma_{\bar c}}_{_\Box}(T,q) = P^{\Gamma_c}_{_\Box}(T^{-1},q^{-1})
\label{blackwhite}
\ee
(in colored superpolynomials this $Z_2$ mirror-like transform also
transpones  the Young diagram, which describes the representation).

Differentials $d_i$ are linear combinations
(actually, sums with plus and minus signs)
of elementary linear maps, associated with particular edges of the hypercube.
We already now the operators, associated with elementary flips
in the cut-and-join operator (\ref{Wop}).
Now we should promote them to linear maps, acting between the tensor
products of vector spaces $V$.

\section{\!\!\!step. Specifying the differentials}

\subsection{The basic maps}

Since every elementary map is associated with some flip of the resolutions,
which splits one of the cycles $\gamma_a$ into two $\gamma_b \cob \gamma_c$
or glues two into one, the map should act either as
$\ Q:\ V_a \longrightarrow V_b\otimes V_c\ $
or as $\ Q^*:\ V_b\otimes V_c \longrightarrow V_a\ $ respectively.
Choosing the $q$-graded basis $v_\pm$ in each $2$-dimensional $V$,
we can describe the maps again in the form of a cut-and-join operator
\be
\hat w = \underbrace{(v_+\otimes v_- + v_-\otimes v_+)\frac{\p}{\p v_+}
+ v_-\otimes v_-\frac{\p}{\p v_-}}_Q
+\ \underbrace{v_+\frac{\p}{\p v_+}\otimes\frac{\p}{\p v_+} + v_-\left(
\frac{\p}{\p v_+}\otimes\frac{\p}{\p v_-} + \frac{\p}{\p v_-}\otimes\frac{\p}{\p v_+}
\right)}_{Q^*}
\ee
The two constituents $Q$ and $Q^*$ of $W$ can be represented as $4\times 2$
and $2\times 4$ matrices respectively:
\be
\begin{array}{ccccccl}
Q: & & V \longrightarrow V\otimes V &&
\left(\begin{array}{rc} 0 & 0 \\ 1 & 0 \\ 1 & 0 \\ 0 & 1 \end{array}\right)
& = & \left(\begin{array}{cc} Q_{++}^+ & Q_{++}^- \\ Q_{+-}^+ & Q_{+-}^- \\
Q_{-+}^+ & Q_{-+}^- \\ Q_{--}^+ & Q_{--}^- \end{array}\right) \\ \\
Q^*:  & &  V\otimes V \longrightarrow   V & &
\left(\begin{array}{cccc} 1 & 0 & 0 & 0 \\ 0 & 1 & 1 & 0 \end{array}\right)
& = & \left(\begin{array}{cccc} Q_+^{++} & Q_+^{+-} & Q_+^{-+} & Q_+^{--} \\
Q_-^{++} & Q_-^{+-} & Q_-^{-+} & Q_-^{--} \end{array}\right)
\end{array}
\ee
Tensorial notation $Q_{ij}^k$ and $Q_i^{jk}$ for $Q$ and $Q^*$
respectively are useful for many purposes.
In particular, in these terms it is easy to formulate important properties
of $Q^*$ and $Q$: they can be considered as defining commutative
\be
\boxed{
Q^i_{jk} = Q^i_{kj}, \ \ \ \ Q^{ij}_k=Q^{ji}_k
}
\label{co}
\ee
and associative
\be
\boxed{
\phantom.[\check Q_j,\check Q_k] = 0,\ \ \ \ [\check Q^j,\check Q^k] = 0
}
\label{coco}
\ee
algebra.
Here the linear maps $\check Q_i$ and $\check Q^i$ are defined as
$(\check Q^i)^j_k = Q^{ij}_k$ and $(\check Q_i)^j_k = Q^j_{ik}$.
The last property is trivial, because
$\ \check Q^- = \check Q_+ = \left(\begin{array}{cc} 1 & 0 \\ 0 & 1 \end{array}\right)\ $
are unit matrices. The other two also coincide:
$\ \check Q^+ = \check Q_- = \left(\begin{array}{cc} 0 & 0 \\ 1 & 0 \end{array}\right)$.

The shape of operations $Q^*$ and $Q$ is severely restricted by the requirement
that they respect $q$-gradation ${\rm grad}_q(v^\pm) = \pm 1$ --
in fact they diminish it by one:
\be
\begin{array}{ccrcl}
Q^*(v^-\otimes v^-) = 0, &&q^{-1}\cdot q^{-1}& \longrightarrow&  0 \\
Q^*(v^+\otimes v^-) = Q^*(v^-\otimes v^+) = v^-,
&& q^{-1}\cdot q &\longrightarrow &   q^{-1} \\
Q^*(v^+\otimes v^+) = v^+,
&& q\cdot q &\longrightarrow &   q \\ \\
Q(v^+) = v^+\otimes v^-\ +\ v^-\otimes v^+, && q &\longrightarrow &  q^{-1}\cdot q\\
Q(v^-) = v^-\otimes v^-, && q^{-1} &\longrightarrow &  q^{-1}\cdot q^{-1}
\end{array}
\ee
In fact, the factor $q^{|r-r_c|}$, which is explicitly
introduced into the definition of Jones polynomials,
is needed to compensate this gradation drop.

\subsection{Differentials and auxiliary  Koszul complex
\label{Kos}}

As usual, from commuting matrices $\tilde Q^i$, which satisfy (\ref{co}),
we can construct
Koszul-like differentials, acting on the functions of anticommuting
variables $\vartheta_i^a$, $\ \vartheta_i^a\vartheta_j^b + \vartheta_j^b\vartheta_i^a = 0$
with arbitrary set of indices $\{a\}$
(see \cite{ADM} for a review in the general context of {\it non-linear algebra} \cite{nolal}).
We take for this indices the labels of the cycles $\gamma_a$
and form a Koszul counterpart of the cut-and-join operator (\ref{Wop}):
\be
{\Q}^\Gamma = \sum_{\stackrel{a}{b<c}} \epsilon^a_{bc}N^a_{bc} \sum_{i,j,k}\left(
Q^{ij}_k\, \vartheta_a^k
\frac{\p^2}{\p \vartheta_b^i\, \p\vartheta_c^j} +
Q_{ij}^k\, \vartheta_b^i \vartheta_b^j\frac{\p}{\p \vartheta_a^k}
\right)
\ee
The $\vartheta$ product is antisymmetric in $bc$, therefore to count each pair $bc$
once one now sums over $b<c$ instead of dividing by two.
We remind that by definition $N^a_{bc}=0$ for $b=c$.
The sign factor $\epsilon^a_{bc} = \pm$ can be chosen in different ways.
The choice, suggested in \cite{BN},
uses the fact that every item with $N^a_{bc}\neq 0$ is associated with
an edge of a hypercube, which connects two hypercube vertices,
differing by just one digit $\star$ in the notation
$r=[\alpha_1\ldots\alpha_{m-1},\star,\alpha_{m+1}\ldots\alpha_n]$.
If $\star$ stands at the $m$-th place, then
\be
\epsilon^a_{bc} = (-)^{\alpha_1+\ldots+\alpha_{m-1}}
\ee

As in (\ref{Wop}) one can associate particular items in this operator
with linear maps, acting along the edges of the hypercube,
and the choice between the two terms in the sum is dictated
by direction of the arrow along the edge.

Operator ${\Q}$ is not nilpotent ${\Q}^2\neq 0$,
it is rather a sum of a BRST-like operator and its conjugate:
${\Q} = {\cal Q} + \widetilde{\cal Q}$
The nilpotent piece, ${\cal Q}$, can be decomposed
into particular pieces, ${\cal Q} = \sum_I d_I$
with the property $d_{I+1}d_I = 0$,
and these $d_I$ can be used as differentials of the complex ${\cal C}$.
Moreover, different colorings $\Gamma_c$ of the graph $\Gamma$
are associated with different complexes and with different
decompositions of ${\Q} = {\cal Q} + \widetilde{\cal Q}$.
The sum of differentials $d_I$ for every $\Gamma_c$ contains exactly one half
of the items of ${\Q}$ -- since arrows along all the edges of hypercube
point in a definite direction.

Decomposition, associated with $\Gamma^c$ can be easily found,
looking at {\it extended} Jones polynomial ${\cal J}^{\Gamma_c}$.
To find $d_I$ one picks up the items of cut-and-join operator
$\hat W^\Gamma$ in (\ref{Wop}), which convert the terms with $t^{I-1}$
in ${\cal J}^{\Gamma_c}$ into those with $t^I$.
Then the same terms in ${\cal Q}$ form the relevant $d_I$.
In the remaining part of this section we illustrate this rule
by several examples.

From now on we use the condensed notation, suppressing the tensor $Q$:
\be
\hat Q_{bc}^{a\,\downarrow}
=\vt_b\vt_c\frac{\p}{\p\vt_a} \equiv Q_{ij}^k\vt_b^i\vt_c^j\frac{\p}{\p\vt_a^k}
= \theta_b\theta_c\frac{\p}{\p\theta_a}
+ (\theta_b\eta_c+\eta_c\theta_b)\frac{\p}{\p\eta_a}, \nn \\
\hat Q_{a\,\uparrow}^{bc} = \vt_a\frac{\p^2}{\p\vt_b\p\vt_c} \equiv
Q^{ij}_k\vt_a^k\frac{\p^2}{\p\vt_b^i\vt_c^j} =
\theta_a\left(\frac{\p^2}{\p\eta_b\p\theta_c} + \frac{\p^2}{\p\theta_b\p\eta_c}
\right) + \eta_a\frac{\p^2}{\p\eta_b\p\eta_c}
\label{elopsusy}
\ee
where at the r.h.s. we also denoted
$\vartheta^+_a=\eta_a$ and $\vartheta^-_a=\theta_a$.
The properties (\ref{co}) and (\ref{coco}) of $Q$ allow to perform
cyclic  permutations of $\vartheta$ -- and this is sufficient
to check the properties $d_{I+1}d_I$ of the differentials
in all the examples below.
Also in what follows the indices $i,j,k$ will be used for other purposes,
and no longer take values $\pm$.

\subsection{Decomposition into differentials $d_I$ for particular $\Gamma_c$}

It can deserve repeating the whole construction of $d_I$ once again.

Cut-and-join operator is associated with the graph $\Gamma$
(the colors of vertices forgotten):
\be
\hat W^\Gamma =
\sum_{E\in {\rm hypercube}^\Gamma}\ \sum_{\beta \,=\, \uparrow,\downarrow}
\hat w_E^{(\beta)}
\ee
where the sum is over edges $E$ of the hypercube associated with $\Gamma$,
and over their orientations $\beta$.
Elementary operators are of the form
$\hat w_E^{(\downarrow)} = p^2\frac{\p}{\p p}$ and
$\hat w_E^{(\uparrow)} = p\frac{\p^2}{\p p^2}$.
If $\Gamma$ has $n$ vertices, then
the $2^n$ vertices of the hypercube are labeled by
binary numbers $[\alpha_1\ldots\alpha_n]$ and its $2^{n-1}n$ edges --
by $[\alpha_1\ldots\alpha_{m-1}\star\alpha_{m+1}\ldots\alpha_m]$:
this edge connects vertices $[\alpha_1\ldots\alpha_{m-1},0,\alpha_{m+1}\ldots\alpha_m]$
and $[\alpha_1\ldots\alpha_{m-1},1,\alpha_{m+1}\ldots\alpha_m]$.

With edge $E$ we associate a sign factor
\be
\epsilon_E = (-)^{\alpha_1+\ldots+\alpha_{m-1}}
\label{sifa}
\ee
Each elementary operator $\hat w_E^{(\beta)}$ has a "supersymmetric" counterpart,
see (\ref{elopsusy}).
Thus one can construct
\be
{\Q}^\Gamma =
\sum_{E } \sum_{\beta \,=\,  \uparrow,\downarrow} \epsilon_E
\hat Q_E^{(\beta)}
\ee

If coloring $\Gamma_c$ of $\Gamma$ is fixed, this implies a choice of
orientation $\beta^c_E$ for each edge.
This turns an edge into an arrow, which has a {\it tail}\,:
the first of the two vertices, that it connects.
The choice of $\beta^c_E$ splits ${\cal Q}$ into two "conjugate" halves,
we pick one of them,
which already has chances to be nilpotent:
\be
{\Q}^\Gamma = {\cal Q}^{\Gamma_c} + \widetilde{{\cal Q}^{\Gamma_c}},\nn \\
{\cal Q}^{\Gamma_c} = \sum_{E }   \epsilon_E\, \hat Q_E^{(\beta^c_E)}
\ee
Existence of {\it tails} allows to further split it into $n$ items:
\be
{\cal Q}^{\Gamma_c} = \sum_{I=0}^n d_I, \nn\\
\boxed{\ \
d_I^{\,\Gamma_c} = \sum_{E: \ |{\rm tail}_E -r_c|=I}    \epsilon_E\,  \hat Q_E^{(\beta^c_E)}
\ \ }
\label{dIQ}
\ee
These $d_I^{\,\Gamma_c}$ are the differentials, associated with $\Gamma_c$.

However, as we already mentioned,
from practical point of view it is often more convenient to simply
read the set of items (\ref{dIQ}) from the expression for extended Jones polynomial.

\subsection{The case of $q=1$: The polynomial $p(T) = P(T,q)|_{q=1}$}

When $q=1$, it is
enough to calculate the ranks of the matrices $d_I$.
All the rest follows from linear algebra relations:
\be
{\rm Rank} = {\rm dim}({\rm Coker})
\ee
and therefore
\be
{\rm dim}\Big({\rm Ker}(d_I)\Big) = {\rm dim}(C_{I-1}) - {\rm dim}\Big({\rm Coker}(d_I)\Big)
= \underline{{\rm dim}(C_{I-1}) - {\rm Rank}(d_I)}
= {\rm dim}\Big({\rm Corank}(d_I)\Big), \nn \\
{\rm dim}\Big({\rm Im}(d_I)\Big) = {\rm dim}\Big({\rm Coker}(\tilde d_I)\Big)
= {\rm Rank}(\tilde d_I) = \underline{{\rm Rank}(d_I)}
= {\rm dim}(C_{I}) - {\rm dim}\Big({\rm Corank}(\tilde d_I)\Big), \nn \\
{\rm dim}\Big({\rm Ker}(d_I)\Big) + {\rm dim}\Big({\rm Im}(d_I)\Big) = {\rm dim}(C_{I-1})
\ee
so that
\be
\!\!p\,(T) =  \sum_{I=0}^n\, T^{I-n_\circ} \left\{{\rm dim}\Big({\rm Ker}(d_{I+1})\Big)
- {\rm dim}\Big({\rm Im}(d_{I})\Big)\right\} =
\sum_{I=0}^n\, T^{I-n_\circ} \Big\{{\rm dim}(C_I) -{\rm Rank}(d_{I+1})
- {\rm Rank}(d_I)\Big\}
\label{p(T)}
\ee
where for the two boundary operators we have ${\rm Rank}(d_{0}) = {\rm Rank}(d_{n+1}) = 0$.

\bigskip

In order to restore the $q$-dependence we need not just
{\it dimensions} of the cohomologies, but $q$-dimensions,
involving the $q$-grading numbers of all their constituents:
\be
P(T,q) =  \frac{q^{n_\bullet}}{(q^2T)^{n_\circ}}
\sum_{I=0}^n\, (qT)^{I} \left\{{\rm dim}_q\Big({\rm Ker}(d_{I+1})\Big)
- {\rm dim}_q\Big({\rm Im}(d_{I})\Big)\right\} = \nn \\
=\frac{q^{n_\bullet}}{(q^2T)^{n_\circ}}
\sum_{I=0}^n\, (qT)^{I} \Big\{{\rm dim}_q(C_I) -{\rm Rank}_q(d_{I+1})
- q^{-1}{\rm Rank}_q(d_I)\Big\} = \nn \\
=\frac{q^{n_\bullet}}{(q^2T)^{n_\circ}}\left(
\sum_{I=0}^n (qT)^I {\rm dim}_q(C_I) - (1+T)\sum_{I=1}^n (qT)^{I-1}{\rm Rank}_q(d_{I})
\right) = \nn \\
=\frac{q^{n_\bullet}}{(q^2T)^{n_\circ}}\left(
(1+T)\sum_{I=1}^n (qT)^{I-1}{\rm dim}_q\Big({\rm Ker}(d_{I})\Big)
-T\sum_{I=0}^{n-1} (qT)^I {\rm dim}_q(C_I) \ \ + \ \ (qT)^n{\rm dim}_q(C_n)\right)
\label{P(T,q)}
\ee
In practice this means that the basis vectors
of $H_I = {\rm Ker}(d_{I+1})/{\rm Im}(d_{I})$ should be found explicitly.
Sometime, technically more convenient is to look for the zero-modes
of Laplace operators $\tilde d_{I+1} d_{I+1}$ and $d_{I}\tilde d_{I}$
perhaps, even among all their eigenvectors
(in the framework of the present
paper all vector spaces are finite-dimensional).

From the last two lines of (\ref{P(T,q)}) it is clear that for $T=-1$
the superpolynomial turns   back to Jones (\ref{Jopo})
\be
P(T,q)\Big|_{T=-1} = J(q)
\ee
The $q$-Euler characteristic of a complex can be calculated both
from $q$-dimensions of cohomologies and of entire spaces $C_I$.

\section{\!\!\!step. Calculating cohomologies}

Now it remains to find the ranks, kernels and images of operators $d_I$.
We continue with the example of $2$-strand braids with $n$ black vertices.
We make just a minor, but important deviation in sec.\ref{dei} to consider
a very simple non-braid example.

\begin{picture}(300,100)(-100,-60)
\put(-20,0){\circle{40}}
\put(20,0){\circle{40}}
\put(0,0){\circle*{5}}
\put(220,0){\circle{40}}
\put(260,0){\circle{40}}
\put(300,0){\circle{40}}
\put(240,0){\circle*{5}}
\put(280,0){\circle*{5}}
\put(-10,-33){\mbox{eight}}
\put(230,-35){\mbox{double eight}}
\put(-13,-43){\mbox{sec.\ref{ei}}}
\put(238,-45){\mbox{sec.\ref{dei}}}
\put(120,0){\circle{30}}
\put(121,16){\circle*{5}}
\put(121,-16){\circle*{5}}
\qbezier(90,0)(90,40)(121,16)
\qbezier(90,0)(90,-40)(121,-16)
\qbezier(150,0)(150,40)(121,16)
\qbezier(150,0)(150,-40)(121,-16)
\put(110,-35){\mbox{Hopf}}
\put(107,-45){\mbox{sec.\ref{hop}}}
\end{picture}

\subsection{Example. An eight: two loops intersecting at one vertex ($n=1$)
\label{ei}}

Extended Jones polynomials in this case are
${\cal J}^\bullet = p_1p_1'+tp_2$ and
${\cal J}^\circ = p_2 + tp_1p_1'$.
Thus the differentials in the two cases look like
\be
d_1^{\,\bullet} = \vt_2\frac{\p^2}{\p\vt_1\p\vt_1'}
=\theta_2\left(\frac{\p^2}{\p\eta_1\p\theta_1'} + \frac{\p^2}{\p\theta_1\p\eta_1'}
\right) + \eta_2\frac{\p^2}{\p\eta_1\p\eta_1'}
= \hat Q_{2\,\uparrow}^{11'}
\ee
and
\be
d_1^{\,\circ} = \vt_1\vt_1'\frac{\p}{\p\vt_2} =
\theta_1\theta_1'\frac{\p}{\p\theta_2}
+ (\theta_1\eta_1'+\eta_1\theta_1')\frac{\p}{\p\eta_2}
= \hat Q_{11'}^{2\,\downarrow}
\ee
i.e. exactly the operators from (\ref{elopsusy}).
These two operators are nilpotent and conjugate to each other, thus the square
${\cal Q} = d_1^{\,\bullet} + d_1^{\,\circ}$ is a non-vanishing "Hamiltonian".

Kernels of these operators depend very much on the space, where they act.
If these spaces are arbitrary functions of Grassmannian variables
$\theta_1,\theta_{1}',\theta_2,\eta_1,\eta_{1}',\eta_2$
the kernels are huge.
However, we are interested in concrete spaces: 4-dimensional $V\otimes V$,
spanned by bilinear functions of $\vartheta_1$ and $\vartheta_1'$
and 2-dimensional $V$ spanned by homogeneous linear function of $\vartheta_2$.

\subsubsection{The $\bullet$ case}

The map
$$
d_1^{\,\bullet}: \ \ V\otimes V \longrightarrow V,
$$
\vspace{-0.5cm}
\be
d_1^{\,\bullet} \Big(\underbrace{\theta_1\theta_1'}_{q^{-2}},\
\theta_1\eta_1'+\eta_1\theta_1',
\ \underbrace{\theta_1\eta_1'-\eta_1\theta_1'}_{q^0},\ \eta_1\eta_1'\Big)
= \Big(0, \  -2\theta_2 ,\ 0,\ -\eta_2\Big)
\ee
has rank two, 2-dimensional kernel and vanishing coimage.
Thus
\be
p^\bullet(T) \ \stackrel{(\ref{p(T)})}{=}\
\Big\{{\rm dim}(C_0) -{\rm Rank}(d_{1}) \Big\} +
T\Big\{{\rm dim}(C_1) -{\rm Rank}(d_{1}) \Big\} =
(4-2) + T(2-2) = 2; \nn \\
P^\bullet(T,q) = q\left({\rm dim}_q\Big({\rm Ker}(d_1)\Big)
+ qT{\rm dim}_q\Big({\rm Coim}(d_1)\Big)\right)
= q(q^{-2}+1) = D
\ee

\subsubsection{The $\circ$ case}

At the same time, the map
$$
d_1^{\,\circ}: \ \ V \longrightarrow V\otimes V,
$$
\vspace{-0.5cm}
\be
d_1^{\,\circ}\Big(\theta_2,\ \eta_2\Big) =
\Big( \theta_1\theta_1' ,\ (\theta_1\eta_1'+\eta_1\theta_1') \Big)
\ee
also has rank two, but vanishing kernel and 2-dimensional coimage,
with the basis $\Big(\underbrace{(\theta_1\eta_1'+\eta_1\theta_1')}_{q^0}, \
\underbrace{\eta_1\eta_1'}_{q^2}\Big)$.
Thus
\be
p^\circ(T) \ \stackrel{(\ref{p(T)})}{=}\
T^{-1}\left(\Big\{{\rm dim}(C_0) -{\rm Rank}(d_{1}) \Big\} +
T\Big\{{\rm dim}(C_1) -{\rm Rank}(d_{1}) \Big\}\right) =
(2-2) + T(4-2) = 2; \nn \\
P^\bullet(T,q) = \frac{1}{q^2T}\left({\rm dim}_q\Big({\rm Ker}(d_1)\Big)
+ qT{\rm dim}_q\Big({\rm Coim}(d_1)\Big)\right)
= \frac{qT}{q^2T}(q^{2}+1) = D
\label{Punknot}
\ee

\subsection{Example. Hopf link ($n=2$)
\label{hop}}

In this case we have four different cycles of length $2$ and two cycles of length $4$.
Extended Jones polynomials are
\be
{\cal J}^{\bullet\bullet} = \bar p_2\bar p_2' + t(p_4+p_4') + t^2p_2p_2', \nn \\
{\cal J}^{\bullet\circ} = p_4+ t(\bar p_2\bar p_2' + p_2p_2') + t^2p_4', \nn \\
{\cal J}^{\circ\bullet} = p_4'+ t(\bar p_2\bar p_2' + p_2p_2') + t^2p_4, \nn \\
{\cal J}^{\circ\circ} =  p_2p_2' + t(p_4+p_4') + t^2\bar p_2\bar p_2'
\ee

\subsubsection{The $\bullet\bullet$ case}

This means that in the first case the differentials are:
\be
d_1^{\bullet\bullet} = (\vt_4+\vt_4')\frac{\p^2}{\p\bar\vt_2\p\bar\vt_2'}
= (\theta_4+\theta_4')\left(\frac{\p^2}{\p\eta_2\p\theta_2'} + \frac{\p^2}{\p\theta_2\p\eta_2'}
\right) + (\eta_4+\eta_4')\frac{\p^2}{\p\eta_2\p\eta_2'},\nn \\
d_2^{\bullet\bullet} = \vt_2\vt_2'\left(\frac{\p}{\p\vt_4} - \frac{\p}{\p\vt_4'}\right)
= \theta_2\theta_2'\left(\frac{\p}{\p\theta_4} - \frac{\p}{\p\theta_4'}\right)
+ (\theta_2\eta_2'+\eta_2\theta_2')\left(\frac{\p}{\p\eta_4}-\frac{\p}{\p\eta_4'}\right)
\ee
\be
{\rm Ker}(d_1^{\bullet\bullet}) = \Big\{\theta_2\theta_2',\
\eta_2\theta_2'-\theta_2\eta_2' \Big\}, \nn \\
{\rm Im}(d_1^{\bullet\bullet}) = \Big\{\theta_4+\theta_4', \
\eta_4+\eta_4'\Big\}  = {\rm Ker}(d_2^{\bullet\bullet}), \nn \\
{\rm Im}(d_2^{\bullet\bullet}) = \Big\{\theta_2\theta_2',\
\theta_2\eta_2'+\eta_2\theta_2'\Big\}
\Longrightarrow \ {\rm Coim}(d_2^{\bullet\bullet}) = \Big\{\eta_2\eta_2',\
\theta_2\eta_2'-\eta_2\theta_2'\Big\}
\ee
and
$$
p^{\bullet\bullet}(T) \ \stackrel{(\ref{p(T)})}{=}\
\Big\{{\rm dim}(C_0) -{\rm Rank}(d_{1}) \Big\} +
T\Big\{{\rm dim}(C_1) -{\rm Rank}(d_{1}) -{\rm Rank}(d_{2})\Big\}
+ T^2\Big\{{\rm dim}(C_2) -{\rm Rank}(d_{2}) \Big\} =
$$
\vspace{-0.6cm}
\be
= (2^2-2)T^0 + (2+2-2-2)T^1 + (2^2-2)T^2 = 2(1+T^2),
\ee
\be
P^{\bullet\bullet}(T,q) = q^2\Big(T^0{\rm dim}_q(H_0^{\bullet\bullet})
+(qT)^1{\rm dim}_q(H_1^{\bullet\bullet})+(qT)^2{\rm dim}_q(H_2^{\bullet\bullet})\Big)
= \nn \\
= q^2\left\{T^0{\rm dim}_q{\rm Ker}(d_1^{\bullet\bullet}) +
(qT)^1\Big({\rm dim}_q{\rm Ker}(d_2^{\bullet\bullet})
-{\rm dim}_q{\rm Im}(d_1^{\bullet\bullet})\Big)
+ (qT)^2{\rm dim}_q{\rm  Coim}(d_2^{\bullet\bullet})\right\} = \nn \\
= q^2\Big\{T^0(q^{-2}+1) + (qT)^1\cdot 0 + (qT)^2(q^{2}+1)\Big\}
= qD(1+q^4T^2)
\label{Pbb}
\ee

\subsubsection{The $\circ\circ$ case}

In the particular case of Hopf link (of $n=2$) the problem with two white
vertices is literally the same as with two black ones, because there
is an (accidental) symmetry $(\bar\vt_2,\bar\vt_2') \leftrightarrow
(\vt_2,\vt_2')$. The only difference in this case when one
switches from black to white is the substitution of overall factor
$q^2$ by $(q^2T)^{-2}$, so that
\be
P^{\circ\circ}(T,q) = (q^2T)^{-2}\Big(q^{-2}P^{\bullet\bullet}(T,q)\Big) =
q^{-1}D(1+q^{-4}T^{-2}) = P^{\bullet\bullet}(T^{-1},q^{-1})
\ee
All intermediate steps in this formula are special for $n=2$,
but the final relation (\ref{blackwhite}) is universal.
However, already for $2$-strand braids with $n>2$ the cohomology
calculus at two sides of the equality will be essentially different,
see s.\ref{Tref} below.

\subsubsection{The $\bullet\circ$ case}

The case of two vertices of different color is different already
for $n=2$.
\be
d_1^{\bullet\circ} = (\vt_2\vt_2'+\bar\vt_2\bar\vt_2')\frac{\p}{\p\vt_4}
= (\theta_2\theta_2'+\bar\theta_2\bar\theta_2')\frac{\p}{\p\theta_4}
+ (\theta_2\eta_2'+\eta_2\theta_2' + \bar\theta_2\bar\eta_2'+\bar\eta_2\bar\theta_2')
\frac{\p}{\p\eta_4}, \nn \\
d_2^{\bullet\circ} = \vt_4'\left(\frac{\p^2}{\p\vt_2\p\vt_2'} -
\frac{\p^2}{\p\bar\vt_2\p\bar\vt_2'}\right) =
\theta_4'\left(\frac{\p^2}{\p\eta_2\p\theta_2'} + \frac{\p^2}{\p\theta_2\p\eta_2'}
-\frac{\p^2}{\p\bar\eta_2\p\bar\theta_2'} - \frac{\p^2}{\p\bar\theta_2\p\bar\eta_2'}\right)
+  \eta_4' \left(\frac{\p^2}{\p\eta_2\p\eta_2'}-\frac{\p^2}{\p\bar\eta_2\p\bar\eta_2'}\right)
\ee
This time
\be
{\rm Ker}(d_1^{\bullet\circ}) = \emptyset
\ \Longrightarrow \ {\rm Corank}(d_1)=0,\ {\rm Rank}(d_1)=2-0=2, \nn \\
{\rm Im}(d_1^{\bullet\circ}) = \Big\{\theta_2\theta_2'+\bar\theta_2\bar\theta_2', \
\theta_2\eta_2'+\eta_2\theta_2' + \bar\theta_2\bar\eta_2'+\bar\eta_2\bar\theta_2'\Big\}, \nn \\
{\rm Coker}(d_2^{\bullet\circ}) =
\Big\{\theta_2\eta_2'+\eta_2\theta_2' - \bar\theta_2\bar\eta_2'-\bar\eta_2\bar\theta_2',
\ \eta_2\eta_2'-\bar\eta_2\bar\eta_2'\Big\}, \nn \\
{\rm Im}(d_2^{\bullet\circ}) = \Big\{\theta_4',\ \eta_4' \Big\}
\Longrightarrow \ {\rm Coim}(d_2^{\bullet\bullet}) = \emptyset,\ \ \
{\rm Rank}(d_2)=2
\ee
so that

\centerline{$
p^{\bullet\circ}(T) \ \stackrel{(\ref{p(T)})}{=}\
T^{-1}\left(\Big\{{\rm dim}(C_0) -{\rm Rank}(d_{1}) \Big\} +
T\Big\{{\rm dim}(C_1) -{\rm Rank}(d_{1}) -{\rm Rank}(d_{2})\Big\}
+ T^2\Big\{{\rm dim}(C_2) -{\rm Rank}(d_{2}) \Big\}\right) =
$}
\vspace{-0.4cm}
\be
= T^{-1}\Big((2-(2-0))T^0 + (2^2+2^2-2-2)T^1 + (2-2)T^2\Big) = 4,
\ee
\be
P^{\bullet\circ}(T,q) =
\frac{q}{q^2T}\Big(T^0{\rm dim}_q(H_0^{\bullet\circ})
+(qT)^1{\rm dim}_q(H_1^{\bullet\circ})+(qT)^2{\rm dim}_q(H_2^{\bullet\circ})\Big)
= \nn \\
= \frac{q}{q^2T}\left\{T^0{\rm dim}_q{\rm Ker}(d_1^{\bullet\circ}) +
(qT)^1\Big({\rm dim}_q{\rm Ker}(d_2^{\bullet\circ})
-{\rm dim}_q{\rm Im}(d_1^{\bullet\circ})\Big)
+ (qT)^2{\rm dim}_q{\rm  Coim}(d_2^{\bullet\circ})\right\} = \nn \\
= (qT)^{-1}\left(T^0\cdot 0 + (qT)^1\Big( (2D^2 - q^2-1) - (q^{-2}+1) \Big)
+ (qT)^2\cdot 0\right) = D^2 = \Big( P^{\rm uknot}(q,T)\Big)^2
\ee
as it should be.

The case $\circ\bullet$ is absolutely the same.

\subsection{Example. Double eight ($n=2$)
\label{dei}}

In fact for $n=2$ there is another knot diagram: "double eight",
which is not a braid.
This knot is topologically trivial for all colorings of the two vertices,
still its consideration is quite instructive.

Four double eight we have four cycles:
two complementary pairs of lengths $1$ and $4$, one cycle of length $2$ and
one of length 5. The hypercube in this case is the square
$\ \ \begin{array}{ccccc}
&&&&\\
&&p\,_1p\,_4\!\!\!&&\\
&\nearrow&\ [10]&\searrow&\\
p\,_1p\,_2p\,_1'\!\!\! &&&& \!\!\! p\,_5\\
\phantom.[00] &\searrow&&\nearrow&\!\!\![11]\\
&&p\,_4'p\,_1'\!\!\!&&\\
&&\ [01]&&
\end{array}.\ \ $

\subsubsection{The $\bullet\bullet$ case}

The choice of arrows is shown for the case two black vertices, when
\be
{\cal J}^{\bullet\bullet} = p\,_1p\,_2p\,_1' + t (p\,_1p\,_4+p\,_1'p\,_4') + t^2p\,_5
\ee

The two differentials are
$$
\!\!
d_1^{\bullet\bullet}
= \vt_4 \frac{\p^2}{\p \vt_2\p \vt_1'} + \vt_4'\frac{\p^2}{\p \vt_2\p \vt_1}
=\theta_4\left(\frac{\p^2}{\p\eta_2\p\theta_1'} + \frac{\p^2}{\p\theta_2\p\eta_1'}
\right) + \eta_4\frac{\p^2}{\p\eta_2\p\eta_1'}
+ \theta_4'\left(\frac{\p^2}{\p\eta_2\p\theta_1} + \frac{\p^2}{\p\theta_2\p\eta_1}
\right) + \eta_4'\frac{\p^2}{\p\eta_2\p\eta_1},
$$
\vspace{-0.4cm}
\be
\!\!
d_2^{\bullet\bullet} = \vt_5\left(\frac{\p^2}{\p\vt_1\p\vt_4} + \frac{\p^2}{\p\vt_1'\p\vt_4'}\right)
=\theta_5\left(\frac{\p^2}{\p\eta_1\p\theta_4} + \frac{\p^2}{\p\theta_1\p\eta_4}
+ \frac{\p^2}{\p\eta_1'\p\theta_4'} + \frac{\p^2}{\p\theta_1'\p\eta_4'}\right)
+ \eta_5\left(\frac{\p^2}{\p\eta_1\p\eta_4} + \frac{\p^2}{\p\eta_1'\p\eta_4'}\right)
\ee
Clearly, $d_2^{\bullet\bullet}d_1^{\bullet\bullet}=0$ and
\be
{\rm Ker}(d_1^{\bullet\bullet}) = \Big\{ \theta_1\theta_2\theta_1',\ \ \theta_1\theta_2\eta_1'
-\theta_1\eta_2\theta_1' + \eta_1\theta_2\theta_1' \Big\},
\ \ \ \ {\rm dim}_q{\rm Ker}(d_1) = q^{-3} + q^{-1}, \nn \\
{\rm Coim}(d_1^{\bullet\bullet}) =
{\rm Coker}(d_2^{\bullet\bullet})
= \Big\{\eta_1\theta_4+\theta_1\eta_4+\eta_1'\theta_4'+\theta_1'\eta_4',
\ \ \eta_1\eta_4+\eta_1'\eta_4'\Big\}, \nn \\
{\rm Im}(d_2^{\bullet\bullet}) = \Big\{\, \theta_5, \  \eta_5\Big\}, \ \ \ \
{\rm Coim}(d_2^{\bullet\bullet}) = \emptyset
\label{deikers}
\ee
At $q=1$, since
\be
{\rm dim}(C_0)={\rm dim}(C_1) =8,\ \ {\rm dim}(C_2)=2\ \ \
{\rm and} \ \ \ {\rm dim}\Big({\rm Rank}(d_1^{\bullet\bullet})\Big)=8-2=6, \ \
{\rm dim}\Big({\rm Rank}(d_2^{\bullet\bullet})\Big) = 2
\ee
we have
\be
p_{de}^{\bullet\bullet}(T) \ \stackrel{(\ref{p(T)})}{=} \
T^0(8-6) + T^1(8-6-2) + T^2(2-2) = 2
\ee
and from the knowledge of $q$-dimensions in (\ref{deikers}) we get:
\be
P_{de}^{\bullet\bullet}(T,q) =
q^2\Big(T^0(q^{-3}+q^{-1}) + (qT)^1\cdot 0 + (qT)^2\cdot 0\Big) = q^{-1}+q = D
\label{Pdebb}
\ee
i.e. coincides with the the answer (\ref{Punknot}) for the unknot,
as necessary.

\bigskip

One can now deduce the same answer for the double-eight knot
with three other colorings of vertices (black-white, white-black and white-white).

\subsubsection{The $\bullet\circ$ case
\label{debw}}

Most interesting is the black-white case with
\be
{\cal J}^{\bullet\circ} = p\,_1p\,_4 + t (p\,_1p\,_2p\,_1'+p\,_5) + t^2\,p_1'p\,_4'
\ee
This is our first example, when differentials are not homogeneous:
\be
d_1^{\bullet\circ} = \vt_2\vt_1'\frac{\p}{\p\vt_4} +\vt_5\frac{\p^2}{\p \vt_1\p\vt_4}
= \theta_2\theta_1' \frac{\p}{\p\theta_4}
+ (\theta_2\eta_1'+\eta_2\theta_1') \frac{\p}{\p\eta_4}
+\theta_5\left(\frac{\p^2}{\p\eta_1\p\theta_4} + \frac{\p^2}{\p\theta_1\p\eta_4}\right)
+ \eta_5\frac{\p^2}{\p\eta_1\p\eta_4},\nn \\
d_2^{\bullet\circ} = \vt_4'\frac{\p^2}{\p \vt_1\p\vt_2} +
\vt_1'\vt_4'\frac{\p}{\p \vt_5} =
\theta_4'\left(\frac{\p^2}{\p\eta_1\p\theta_2} + \frac{\p^2}{\p\theta_1\p\eta_2}\right)
+ \eta_4'\frac{\p^2}{\p\eta_1\p\eta_2}
 + \theta_1'\theta_4' \frac{\p}{\p\theta_5}+
(\theta_1'\eta_4'+\eta_1'\theta_4') \frac{\p}{\p\eta_5}
\ee
Again, $d_2^{\bullet\circ}d_1^{\bullet\circ}=0$,
but kernels, images and even cohomologies are quite different from (\ref{deikers}):
\be
{\rm Ker}(d_1^{\bullet\circ}) = \emptyset, \ \ \
\Longrightarrow\ {\rm Rank}(d_1^{\bullet\circ}) = 4  \nn \\
{\rm Im}(d_1^{\bullet\circ}) =  \Big\{\theta_1\theta_2\theta_1',\ \
(\theta_1\eta_2+\eta_1\theta_2)\theta_1' + \theta_1\theta_2\eta_1',\
\ \theta_5-\eta_1\theta_2\theta_1',\ \
\eta_5-\eta_1(\theta_2\eta_1'+\eta_2\theta_1')\Big\},\nn \\
{\rm Coker}(d_2^{\bullet\circ}) = {\rm Im}(d_1^{\bullet\circ})
\oplus \Big\{ (\theta_1\eta_2-\eta_1\theta_2)\theta_1',\ \
(\theta_1\eta_2-\eta_1\theta_2)\eta_1' \Big\}, \nn \\
{\rm Coim}(d_2^{\bullet\circ}) = \emptyset, \ \ \
\Longrightarrow\ {\rm Rank}(d_2^{\bullet\circ}) = 4
\label{deikersbw}
\ee
(note, that basis elements are not obligatory homogeneous).
Thus
\be
p^{\bullet\circ}_{de}(T) = T^{-1}\Big( T^0(2^2-4) + T^1(2^3+2-4-4) + T^2(4-4)\Big)=2,\nn \\
P^{\bullet\circ}_{de}(T,q) = \frac{q}{q^2T}\Big(T^0\cdot 0
+ (qT)( q+q^{-1})
+ (qT)^2\cdot 0\Big) = D
\ee
in accordance with (\ref{Pdebb}).

\bigskip

The white-black case is fully symmetric to black-white and does not bring
anything new.

\subsubsection{The $\circ\circ$ case}

The white-white case, however, is not identical to black-black.
Unlike it happened with the Hopf link in sec.\ref{hop},
the white-white cohomologies are now different from black-black
(in a trivial way, of course, since the example is trivial)
-- only superpolynomial remains the same.
Extended Jones polynomial
\be
{\cal J}^{\bullet\bullet} =  p\,_5+ t (p\,_1p\,_4+p\,_1'p\,_4') + t^2p\,_1p\,_2p\,_1'
\ee
implies that
\be
d_1^{\circ\circ} = (\vt_1\vt_4+\vt_1'\vt_4')\frac{\p}{\p\vt_5}
=(\theta_1\theta_4+\theta_1'\theta_4') \frac{\p}{\p\theta_5}
+(\theta_1\eta_1+\eta_1\theta_1+\theta_1'\eta_1'+\eta_1'\theta_1')
\frac{\p}{\p\eta_5},\nn\\
d_2^{\circ\circ} = \vt_1\vt_2 \frac{\p^2}{\p\vt_4'} - \vt_1'\vt_2\frac{\p}{\p \vt_4}
= \theta_1\theta_2 \frac{\p}{\p\theta_4'}
+ (\theta_1\eta_2+\eta_1\theta_2) \frac{\p}{\p\eta_4'}
- \theta_1'\theta_2 \frac{\p}{\p\theta_4}
- (\theta_1'\eta_2+\eta_1'\theta_2) \frac{\p}{\p\eta_4}
\ee
Thus
\be
{\rm Ker}(d_1^{\circ\circ}) = \emptyset, \ \ \
\Longrightarrow\ {\rm Rank}(d_1^{\bullet\circ}) = 2  \nn \\
{\rm Im}(d_1^{\bullet\circ})
= {\rm Ker}(d_2^{\bullet\circ})\ \ \
\Longrightarrow\ {\rm Rank}(d_2^{\bullet\circ}) = 6\nn \\
{\rm Coim}(d_2^{\bullet\circ}) = \Big\{ -\theta_1\eta_2\eta_1'
+ 2\eta_1\theta_2\eta_1' - \eta_1\eta_2\theta_1',\ \ \eta_1\eta_2\eta_1'\Big\},
\label{deikersww}
\ee
so that
\be
p^{\circ\circ}_{de}(T) = T^{-2}\Big( T^0(2-2) + T^1(2^2+2^2-2-6) + T^2(2^3-6)\Big)=2,\nn \\
P^{\circ\circ}_{de}(T,q) = \frac{1}{(q^2T)^2}\Big(T^0\cdot 0 + (qT)\cdot 0
+ (qT)^2(q + q^3)\Big) = D = P^{\bullet\bullet}_{de}(T,q)  = P^{\rm unknot}(T,q)
\ee
As promissed, this time $H_2^{\circ\circ}\cong V$, while in the "mirror case"
non-vanishing was $H^{\bullet\bullet}_0 \cong V$.

\subsection{Example: Trefoil ($n=3$)
\label{Tref}}

Differentials are immediately written if we use ${\cal Q}$
and formulas (\ref{Jontre}).

\subsubsection{Three black vertices: the starting vertex is $r_c=[000]=p_3p_3'$}

Since, according to (\ref{Jontre}), the extended Jones polynomial in this case is
$${\cal J}^{\bullet\bullet\bullet} = p_3p_3' + t(p_6+p_6'+p_6'') +
t^2(p_4p_2+p_4'p_2'+p_4''p_2'') + t^3p_2p_2'p_2'',$$
we get for the differentials:
\be
d_1 = (\vartheta_6+\vartheta_6'+\vartheta_6'')\frac{\p^2}{\p\vartheta_3\p\vartheta_3'},\nn\\
d_2 = (\vt_4'\vt_2' - \vt_4''\vt_2'')\frac{\p}{\p\vt_6}
+ (\vt_4''\vt_2''-\vt_4\vt_2)\frac{\p}{\p\vt_6'}
+ (\vt_4\vt_2-\vt_4'\vt_2')\frac{\p}{\p\vt_6''}, \nn \\
d_3 = \vt_2\vt_2'\frac{\p}{\p\vt_4''} + \vt_2''\vt_2\frac{\p}{\p\vt_4'}
+ \vt_2'\vt_2''\frac{\p}{\p\vt_4}
\ee
Convolution with $Q$-tensors is implied, but suppressed
(i.e. actually
$\vt_a\vt_b\frac{\p}{\p\vt_c} = Q_{ij}^k\vt_a^i\vt_b^j\frac{\p}{\p\vt_c^k}$),
the properties of $Q$
allow cyclic permutations of $\theta$'s. and this is all what is needed to prove
that
\be
d_3d_2=d_2d_1=0
\ee

We do not consider cohomologies in this case, because this will be done
in more generality in s.\ref{genblack} below,
and -- by a more primitive method --
in Appendix B at the end of the paper.

\subsubsection{The case $\Gamma_c = \circ\bullet\bullet\ $: $r_c=[100]=p_6$}

Since $${\cal J}^{\circ\bullet\bullet}
= p_6 + t( p_4'p_2' + p_4''p_2'' + p_3p_3') +
t^2(p_6'+p_6''+p_2p_2'p_2'') + t^3p_4p_2,$$ we have:
\be
d_1 = (\vt_4'\vt_2' +\vt_4''\vt_2'' + \vt_3\vt_3')\frac{\p}{\p\vt_6}, \nn \\
d_2 = (\vt_6' + \vt_6'')\frac{\p^2}{\p\vt_3\p\vt_3'}
- \vt_6'\frac{\p^2}{\p\vt_4''\p\vt_2''} - \vt_6''\frac{\p^2}{\p\vt_4'\p\vt_2'}
+ \vt_2\vt_2''\frac{\p}{\p\vt_4'} + \vt_2\vt_2'\frac{\p}{\p\vt_4''}, \nn \\
d_3 = \vt_4\vt_2\left(\frac{\p}{\p\vt_6'}-\frac{\p}{\p\vt_6''}\right)
+ \vt_4\frac{\p^2}{\p\vt_2'\p\vt_2''}
\ee
As in above example (\ref{debw}), the differentials are not homogeneous,
thus their kernel and images have somewhat sophisticated basises.

\subsection{Generic $n$, all vertices black
\label{genblack}}

As in all previous examples of 2-strand braids,
we label the cycles by index $a = (n_a,i)$,
i.e. by its length and additional index $i$,
enumerating different cycles of the same length.

The $2$-strand braid with $n$ vertices is a graph $\Gamma$
with $2n$ edges, which connect pairwise the two subsequent
vertices.
Considering various resolutions of crossings at the vertices,
it is clear that
there are two cycles of length $n$
and $n$ cycles of each even length: $2, 4, 6, \ldots, 2n$,
i.e. there are two time-variables $p\,_n,p\,_n'$ and
also $p\,_{2k}^{(i)}$ with $k=1,\ldots,n$ and $i=1,\ldots,n$
(even if $n$ is even, the cycles, $\gamma_n,\gamma_n'$ are different
from $\gamma_n^{(i)}$ and also different are the time-variables).
If vertices are enumerated in a natural way,
one can think of the cycle $p_{2k}^{(i)}$ as consisting of
all edges of $\Gamma$ between the vertices $i$ and $i+k$.

Then it is clear that the extended Jones polynomial in this case is
\be
{\cal J}^{(n)} = p_np_n' + t\sum_{i=1}^n p_{2n}^{(i)} +
t^2 \sum_{i<j} p^{(i)}_{2(j-i)} p^{(j)}_{2n-2(j-i)} + \ldots
+\ t^k\!\!\!\!\!\! \sum_{i_1<i_2<\ldots <i_k}p^{(i_1)}_{2(i_2-i_1)}
p^{(i_2)}_{2(i_3-i_2)}\ldots p^{(i_k)}_{2n-2(i_k-i_1)} +\nn \\
+ \ldots + t^n \prod_{i=1}^n p_2^{(i)}
\label{extnJon}
\ee

From  (\ref{extnJon}) the ordinary unreduced Jones polynomial is
\be
J^{(n)} = q^n\left(D^2 - nqD+\frac{n(n-1)}{2}q^2D^2 - \ldots\right) =
q^n\Big(D^2-1 + (1-qD)^n\Big) = q^n\Big(q^{-2}+1+q^{2} + (-q^2)^n\Big)
\ee
and consists of four items -- this is a familiar formula, which
we already derived in many ways in this text.

\bigskip

Moat important, from (\ref{extnJon}) we straightforwardly read the differentials.

\subsubsection{Differential $d_1$}

\be
d_1 = \left(\sum_{i=1}^n \vartheta_{2n}^{(i)}\right)\frac{\p^2}{\p\vt_n\p\vt_n'}
\label{d1bbbb}
\ee
As usual for $2$-strand knots, the $d_1$ is a product of two independent
factors, what matters is the second one,
which does not actually depend on $n$.
Thus, as in all previous examples,
the rank of $d_1$ is two, its  kernel and its image are
\be
{\rm Ker}(d_1) = \{\theta_n\theta_n',\ (\theta_n\eta_n'-\eta_n\theta_n')\},
\ \ \ \ {\rm dim}_q(H_0) = {\rm dim}_q{\rm Ker}(d_1) = q^{-2}+1, \nn \\
{\rm Im}(d_1) = \left\{\sum_{i=1}^n\theta_{2n}^{(i)},\ \
\sum_{i=1}^n \eta_{(2n)}^{(i)} \right\}, \ \ \ \
{\rm dim}_q{\rm Im}(d_1) =  q^{-1}+q
= q^{-1}\Big({\rm dim}_q(C_0) -{\rm dim}_q{\rm Ker}(d_1)\Big)
\label{kerimd1n}
\ee
In the last equality the factor $q^{-1}$ appears because our differentials
decrease $q$-grading by one.

\subsubsection{Differential $d_2$}

\be
d_2 =  \sum_{i=1}^n \left(-\sum_{j>i} \vt_{2j-2i}^{(i)}\vt_{2n+2i-2j}^{(j)} \
+\ \sum_{j<i} \vt_{2i-2j}^{(j)}\vt_{2n+2j-2i}^{(i)}\right)
\frac{\p}{\p\vt_{2n}^{(i)}}
\label{d2n}
\ee
The terms with $j>i$ in this sum are associated with the edges
$[0\ldots 0 \stackrel{i}{1} 0\ldots 0\stackrel{j}{\star} 0\ldots 0]$
of the hypercube, connecting its vertices
$[0\ldots 0\stackrel{i}{1}0\ldots 0\stackrel{j}{0} 0\ldots 0]$ (i.e. $p_{2n}^{(i)}$)
and
$[0\ldots 0\stackrel{i}{1} 0\ldots 0\stackrel{j}{1} 0\ldots 0]_{\phantom{\int_\sum}}\!\!\!\!$
(i.e. $p_{2j-2i}^{(i)}p_{2n+2i-2j}^{(j)}$).
Therefore, according to the rule (\ref{sifa}) they enter with the
sign factor $\epsilon_E = -1$.
The terms with $j<i$ are associated with edges
$[0\ldots 0\stackrel{j}{\star} 0\ldots 0\stackrel{i}{1} 0\ldots 0]$
between
$[0\ldots 0\stackrel{j}{0} 0\ldots 0\stackrel{i}{1} 0\ldots 0]$
and
$[0\ldots 0\stackrel{j}{1} 0\ldots 0\stackrel{i}{1} 0\ldots 0]_{\phantom{\int_\sum}}
\!\!\!\!\!\!\!$,
and the sign factor is $+1$. \\

Clearly, $d_2d_1=0$.
The kernel of $d_2$ is non-vanishing only because the sum over
of the bracket over $i$ is zero, thus iff all derivatives
$\frac{\p}{\p\vt_{2n}^{(i)}}$ are the same, the action of $d_2$
gives zero. This means that
\be
{\rm Ker}(d_2)
= \left\{\sum_{i=1}^n \theta^{(i)}_{2n},\ \sum_{i=1}^n \eta^{(i)}_{2n}\right\},
\ \ \ \ {\rm dim}_q{\rm Ker}(d_2) = q^{-1}+q = D
\ee
Comparing with the second line of (\ref{kerimd1n}), we conclude
that ${\rm Ker}(d_2) = {\rm Im}(d_1)$, and
\be
H_1 = {\rm Ker}(d_2)\Big/\,{\rm Im}(d_1) = \emptyset
\ee

As to the image of $d_2$,
\be
{\rm dim}_q{\rm Im}(d_2) = q^{-1}\Big({\rm dim}_q(C_1) - {\rm dim}_q{\rm Ker}(d_2)\Big)
= (n-1)\,q^{-1}D
\label{dimimd2}
\ee

In fact, if we agree to define the cycle lengths modulo $2n$, i.e.
define $\vt^{(i)}_{2k}$ with negative lengths $-2k$ as $\vt^{(i)}_{2n-2k}$,
then $d_2$ acquires a simpler form:
\be
d_2 = \sum_{i,j} \vt^{(i)}_{2j-2i}\vt^{(j)}_{2i-2j}\frac{\p}{\p \vt^{(j)}_{2n}}
= -\sum_{i,j} \vt^{(i)}_{2j-2i}\vt^{(j)}_{2i-2j}\frac{\p}{\p \vt^{(i)}_{2n}}
\label{d2brab}
\ee
(the terms with $i=j$ are automatically excluded by anticommutativity of $\vt$-variables).

\subsubsection{Differential $d_3$}

Pictorially the action of $d_2$ and $d_3$ can be represented as follows:

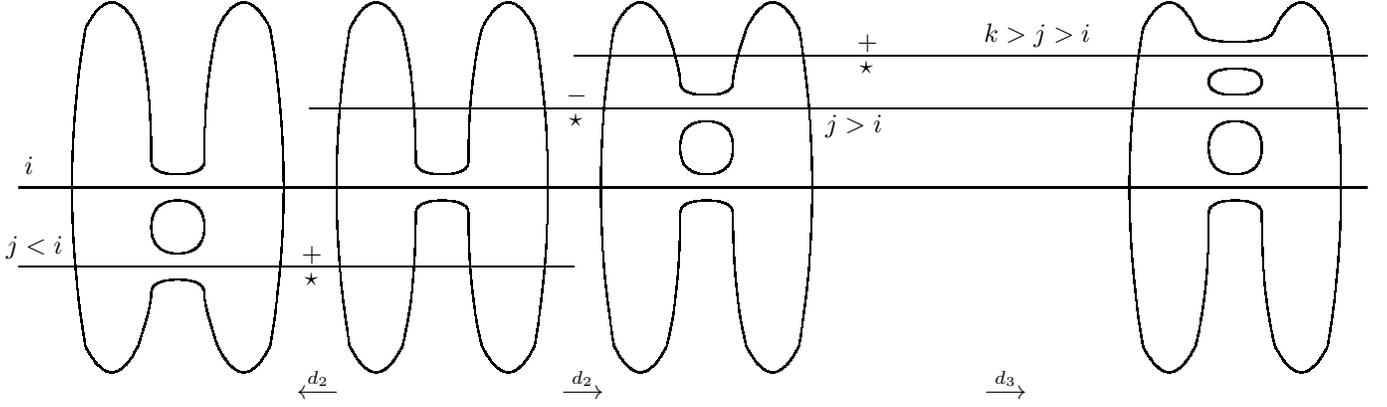
\begin{picture}(300,170)(-135,-85)
\qbezier(-35,-60)(-45,0)(-35,60)
\qbezier(-35,-60)(-25,-80)(-15,-60)
\qbezier(-15,-60)(-10,-40)(-10,-10)
\qbezier(-10,-10)(-10,-5)(0,-5)
\qbezier(-35,60)(-25,80)(-15,60)
\qbezier(-15,60)(-10,40)(-10,10)
\qbezier(-10,10)(-10,5)(0,5)
\qbezier(35,-60)(45,0)(35,60)
\qbezier(35,-60)(25,-80)(15,-60)
\qbezier(15,-60)(10,-40)(10,-10)
\qbezier(10,-10)(10,-5)(0,-5)
\qbezier(35,60)(25,80)(15,60)
\qbezier(15,60)(10,40)(10,10)
\qbezier(10,10)(10,5)(0,5)
\qbezier(65,-60)(55,0)(65,60)
\qbezier(65,-60)(75,-80)(85,-60)
\qbezier(85,-60)(90,-40)(90,-10)
\qbezier(90,-10)(90,-5)(100,-5)
\qbezier(65,60)(75,80)(85,60)
\qbezier(85,60)(90,45)(90,40)
\qbezier(90,15)(90,5)(100,5)
\qbezier(90,15)(90,25)(100,25)
\qbezier(90,40)(90,35)(100,35)
\qbezier(135,-60)(145,0)(135,60)
\qbezier(135,-60)(125,-80)(115,-60)
\qbezier(115,-60)(110,-40)(110,-10)
\qbezier(110,-10)(110,-5)(100,-5)
\qbezier(135,60)(125,80)(115,60)
\qbezier(115,60)(110,45)(110,40)
\qbezier(110,15)(110,5)(100,5)
\qbezier(110,15)(110,25)(100,25)
\qbezier(110,40)(110,35)(100,35)
\qbezier(-65,-60)(-55,0)(-65,60)

\qbezier(-65,-60)(-75,-80)(-85,-60)
\qbezier(-85,-60)(-90,-45)(-90,-40)
\qbezier(-90,-15)(-90,-5)(-100,-5)
\qbezier(-90,-15)(-90,-25)(-100,-25)
\qbezier(-90,-40)(-90,-35)(-100,-35)
\qbezier(-65,60)(-75,80)(-85,60)
\qbezier(-85,60)(-90,40)(-90,10)
\qbezier(-90,10)(-90,5)(-100,5)
\qbezier(-135,-60)(-145,0)(-135,60)
%
\qbezier(-135,-60)(-125,-80)(-115,-60)
\qbezier(-115,-60)(-110,-45)(-110,-40)
\qbezier(-110,-15)(-110,-5)(-100,-5)
\qbezier(-110,-15)(-110,-25)(-100,-25)
\qbezier(-110,-40)(-110,-35)(-100,-35)
\qbezier(-135,60)(-125,80)(-115,60)
\qbezier(-115,60)(-110,40)(-110,10)
\qbezier(-110,10)(-110,5)(-100,5)
\qbezier(265,-60)(255,0)(265,60)
\qbezier(265,-60)(275,-80)(285,-60)
\qbezier(285,-60)(290,-40)(290,-10)
\qbezier(290,-10)(290,-5)(300,-5)
\qbezier(265,60)(275,80)(285,60)
\qbezier(290,15)(290,5)(300,5)
\qbezier(290,15)(290,25)(300,25)
\qbezier(290,40)(290,35)(300,35)
\qbezier(290,40)(290,45)(300,45)
\qbezier(285,60)(286,55)(300,55)
\qbezier(335,-60)(345,0)(335,60)
\qbezier(335,-60)(325,-80)(315,-60)
\qbezier(315,-60)(310,-40)(310,-10)
\qbezier(310,-10)(310,-5)(300,-5)
\qbezier(335,60)(325,80)(315,60)
\qbezier(310,15)(310,5)(300,5)
\qbezier(310,15)(310,25)(300,25)
\qbezier(310,40)(310,35)(300,35)
\qbezier(310,40)(310,45)(300,45)
\qbezier(315,60)(314,55)(300,55)
\put(-160,0){\line(1,0){510}}
\put(-160,-30){\line(1,0){210}}
\put(-50,30){\line(1,0){400}}
\put(-158,5){\mbox{$i$}}
\put(-165,-25){\mbox{$j<i$}}
\put(145,21){\mbox{$j>i$}}
\put(50,50){\line(1,0){300}}
\put(205,55){\mbox{$k>j>i$}}
\put(-52,-37){\mbox{$\star$}}
\put(-53,-28){\mbox{$+$}}
\put(48,23){\mbox{$\star$}}
\put(47,32){\mbox{$-$}}
\put(158,43){\mbox{$\star$}}
\put(157,52){\mbox{$+$}}
\put(-55,-80){\mbox{$\stackrel{d_2}{\longleftarrow}$}}
\put(45,-80){\mbox{$\stackrel{d_2}{\longrightarrow}$}}
\put(205,-80){\mbox{$\stackrel{d_3}{\longrightarrow}$}}
\end{picture}

\noindent
In this picture $d_2$ acts on the cycle $\gamma_{2n}^{(i)}$
of length $2n$, which is
obtained by gluing $\gamma_n$ and $\gamma_n'$ at the vertex $i$.
The action of $d_2$ makes a flip at the vertex $j$,
this splits $\gamma_{2n}^{(i)}$ into two cycles:
one of the length $2(j-i)$ between vertices $i$ and $j$
and a complementary one of the length $2n-2(j-i)$.
The sign factor (\ref{sifa}) depends on the sign $j-i$,
this is also shown in the picture
(of course, the sign depends on the conventions about
the ordering of $\vt$ variables in (\ref{elopsusy})).

Operator $d_3$ acts in just the same way, making a new flip
at vertex $k$.
We show just one case of six possible orderings in the picture: $k>j>i$.
In this case splitted is the cycle $\gamma_{2n-2j+2i}^{(j-i)}$
(complementary to the one between $i$ and $j$), and produced
are the two new cycles of the length $2(k-j)$ and $2n-2(k-i)$.
The sign factor (\ref{sifa}) in this case is positive.

The next picture shows three different elements from the space $C_2$,
which are mapped by $d_3$ into a given element of $C_3$ with $i<j<k\leq n$:

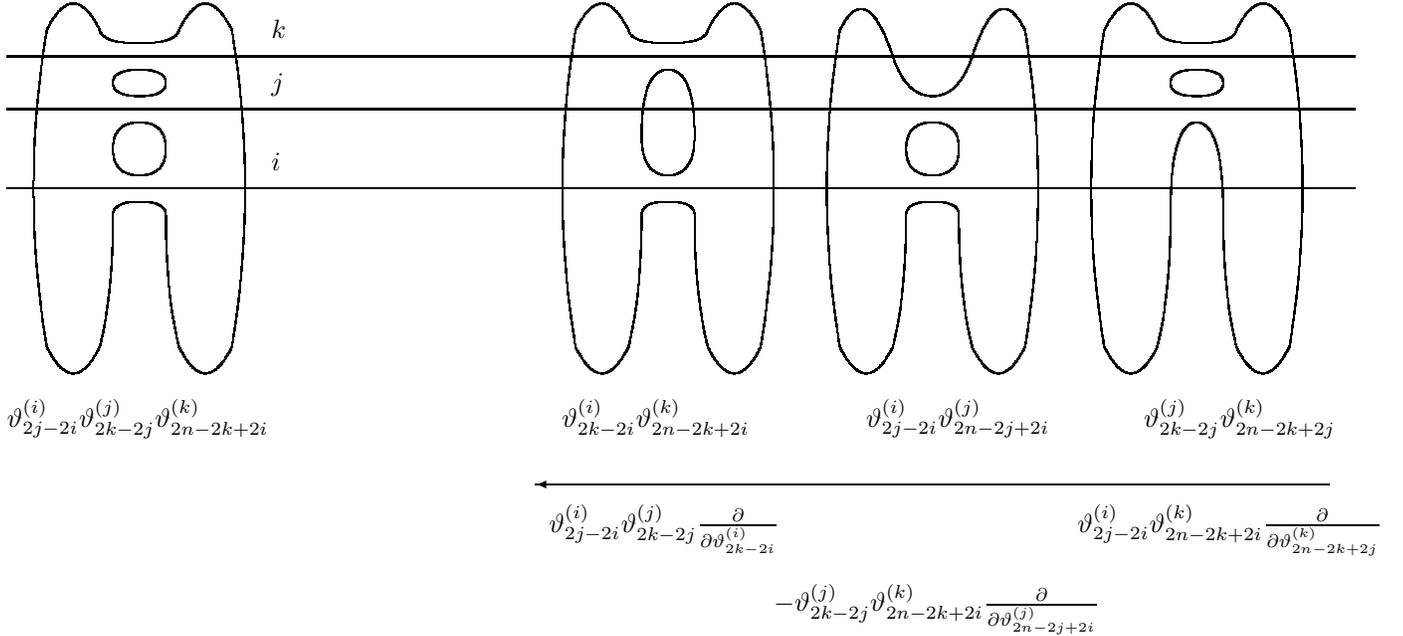
\begin{picture}(300,270)(75,-175)
\qbezier(65,-60)(55,0)(65,60)
\qbezier(65,-60)(75,-80)(85,-60)
\qbezier(85,-60)(90,-40)(90,-10)
\qbezier(90,-10)(90,-5)(100,-5)
\qbezier(65,60)(75,80)(85,60)
\qbezier(90,15)(90,5)(100,5)
\qbezier(90,15)(90,25)(100,25)
\qbezier(90,40)(90,35)(100,35)
\qbezier(90,40)(90,45)(100,45)
\qbezier(85,60)(86,55)(100,55)
\qbezier(135,-60)(145,0)(135,60)
\qbezier(135,-60)(125,-80)(115,-60)
\qbezier(115,-60)(110,-40)(110,-10)
\qbezier(110,-10)(110,-5)(100,-5)
\qbezier(135,60)(125,80)(115,60)
\qbezier(110,15)(110,5)(100,5)
\qbezier(110,15)(110,25)(100,25)
\qbezier(110,40)(110,35)(100,35)
\qbezier(110,40)(110,45)(100,45)
\qbezier(115,60)(114,55)(100,55)
\qbezier(265,-60)(255,0)(265,60)
\qbezier(265,-60)(275,-80)(285,-60)
\qbezier(285,-60)(290,-40)(290,-10)
\qbezier(290,-10)(290,-5)(300,-5)
\qbezier(265,60)(275,80)(285,60)
\qbezier(290,20)(290,5)(300,5)
%
\qbezier(290,20)(290,45)(300,45)
\qbezier(285,60)(286,55)(300,55)
\qbezier(335,-60)(345,0)(335,60)
\qbezier(335,-60)(325,-80)(315,-60)
\qbezier(315,-60)(310,-40)(310,-10)
\qbezier(310,-10)(310,-5)(300,-5)
\qbezier(335,60)(325,80)(315,60)
\qbezier(310,20)(310,5)(300,5)
%
\qbezier(310,20)(310,45)(300,45)
\qbezier(315,60)(314,55)(300,55)
\qbezier(365,-60)(355,0)(365,60)
\qbezier(365,-60)(375,-80)(385,-60)
\qbezier(385,-60)(390,-40)(390,-10)
\qbezier(390,-10)(390,-5)(400,-5)
\qbezier(365,60)(375,80)(385,50)
\qbezier(390,15)(390,5)(400,5)
\qbezier(390,15)(390,25)(400,25)
\qbezier(385,50)(390,35)(400,35)
%
%
%
\qbezier(435,-60)(445,0)(435,60)
\qbezier(435,-60)(425,-80)(415,-60)
\qbezier(415,-60)(410,-40)(410,-10)
\qbezier(410,-10)(410,-5)(400,-5)
\qbezier(435,60)(425,80)(415,50)
\qbezier(410,15)(410,5)(400,5)
\qbezier(410,15)(410,25)(400,25)
\qbezier(415,50)(410,35)(400,35)
%
%
%
%
%
%
%
\qbezier(465,-60)(455,0)(465,60)
\qbezier(465,-60)(475,-80)(485,-60)
\qbezier(485,-60)(490,-40)(490,-10)
\qbezier(465,60)(475,80)(485,60)
\qbezier(490,-10)(490,25)(500,25)
\qbezier(490,40)(490,35)(500,35)
\qbezier(490,40)(490,45)(500,45)
\qbezier(485,60)(486,55)(500,55)
\qbezier(535,-60)(545,0)(535,60)
\qbezier(535,-60)(525,-80)(515,-60)
\qbezier(515,-60)(510,-40)(510,-10)
\qbezier(535,60)(525,80)(515,60)
\qbezier(510,-10)(510,25)(500,25)
\qbezier(510,40)(510,35)(500,35)
\qbezier(510,40)(510,45)(500,45)
\qbezier(515,60)(514,55)(500,55)
\put(50,0){\line(1,0){510}}
\put(50,50){\line(1,0){510}}
\put(50,30){\line(1,0){510}}
\put(150,7){\mbox{$i$}}
\put(150,37){\mbox{$j$}}
\put(150,57){\mbox{$k$}}
\put(50,-90){\mbox{$\vt^{(i)}_{2j-2i}\vt^{(j)}_{2k-2j}\vt^{(k)}_{2n-2k+2i}$}}
\put(260,-90){\mbox{$\vt^{(i)}_{2k-2i}\vt^{(k)}_{2n-2k+2i}$}}
\put(375,-90){\mbox{$\vt^{(i)}_{2j-2i}\vt^{(j)}_{2n-2j+2i}$}}
\put(480,-90){\mbox{$\vt^{(j)}_{2k-2j}\vt^{(k)}_{2n-2k+2j}$}}
\put(255,-130){\mbox{$\vt^{(i)}_{2j-2i}\vt^{(j)}_{2k-2j}\frac{\p}{\p\vt^{(i)}_{2k-2i}}$}}
\put(340,-160){\mbox{$-\vt^{(j)}_{2k-2j}\vt^{(k)}_{2n-2k+2i}\frac{\p}{\p\vt^{(j)}_{2n-2j+2i}}$}}
\put(455,-130){\mbox{$\vt^{(i)}_{2j-2i}\vt^{(k)}_{2n-2k+2i}\frac{\p}{\p\vt^{(k)}_{2n-2k+2j}}$}}
\put(550,-112){\vector(-1,0){300}}
\end{picture}

\noindent
It is easy to see that -- if the cycle lengths are calculated modulo $2n$ --
that all the three cases are described by the single formula
\be
\!d_3 = -\!\sum_{i,j,k}\! \vt^{(i)}_{2j-2i}\vt^{(j)}_{2k-2j}\frac{\p}{\p\vt^{(i)}_{2k-2i}}
\stackrel{(\ref{elopsusy})}{=}
-\!\sum_{i,j,k}\! \left(\theta^{(i)}_{2j-2i}\theta^{(j)}_{2k-2j} \frac{\p}{\p\theta^{(i)}_{2k-2i}}+
\Big(\theta^{(i)}_{2j-2i}\eta^{(j)}_{2k-2j}+\eta^{(i)}_{2j-2i}\theta^{(j)}_{2k-2j}\Big)
\frac{\p}{\p\eta^{(i)}_{2k-2i}}
\right)
\label{d3brab}
\ee

\bigskip

Already at the level of $d_3$ the cohomology calculus is getting rather tedious.

We begin it from the simplest case of $n=3$.
For this we explicitly describe the action of $d_3$ on the $12$ elements of the basis in
$C_2 = V_4\otimes V_2 \oplus V_4'\otimes V_2' \oplus V_4''\otimes V_2''$
(we remind that all vector spaces are isomorphic to two-dimensional $V$,
the labels are, however, important to define the action of $d_3$):
\be
d_3^{\,(n=3)}:\ \ \ \ \ \ \
\begin{array}{cccc|c|c}
\theta^{(i)}_2\theta^{(i+1)}_4 &(i=1,2,3)&\longrightarrow &
\theta^{(1)}_2\theta^{(2)}_2\theta^{(3)}_2 & 3 \longrightarrow 1 & 2q^{-2}\\
\hline
\eta^{(i)}_2\eta^{(i+1)}_4 &(i=1,2,3)&\longrightarrow &
\eta^{(i)}_2\Big(\eta^{(i+1)}_2\theta^{(i+2)}_2 + \theta^{(i+1)}_2\eta^{(i+2)}_2\Big)
& 3 \longrightarrow 3 & 0 \\
\hline
\eta^{(i)}_2\theta^{(i+1)}_4 &(i=1,2,3)&\longrightarrow &
\eta^{(i)}_2\theta^{(i+1)}_2\theta^{(i+2)}_2 & 3 \longrightarrow 3 & 0\\
\theta^{(i)}_2\eta_4^{(i+1)} &(i=1,2,3)&\longrightarrow &
\theta^{(i)}_2\Big(\eta^{(i+1)}_2\theta^{(i+2)}_2+\theta^{(i+1)}_2\eta^{(i+2)}_2\Big)
& 3\ ^\nearrow & 3
\end{array}
\ee
The last column shows the contribution to ${\rm dim}_q{\rm Ker}(d_3)$.
Thus we see that
\be
{\rm for}\ n=3\ \ \ \ {\rm dim}_q{\rm Ker}(d_3) = 2q^{-2} + 3, \nn\\
\Longrightarrow \ \
{\rm dim}_q(H_2^{\,(3)}) = {\rm dim}_q{\rm Ker}(d_3) - {\rm dim}_q{\rm Im}(d_2)
\ \stackrel{(\ref{dimimd2})}{=}\  (2q^{-2}+3) - 2q^{-1}D = 1
\ee
what implies that the contribution of the order $T^2$
to $P^{\bullet\bullet\bullet}(T,q)$ is
$\ q^3(qT)^2\cdot 1 = q^5T^2$.

Next, $n=4$:
\be
d_3^{\,(n=4)}:\ \ \ \ \ \ \
\begin{array}{cccc|c|c}
\theta^{(i)}_2\theta^{(i+1)}_6 &(i=1,\ldots,4)&\longrightarrow &
\theta^{(i)}_2\Big(\theta^{(i+1)}_2\theta^{(i+2)}_4 + \theta^{(i+1)}_4\theta^{(i+3)}_2\Big)
& 4 \longrightarrow 3 & q^{-3}\\
\theta^{(i)}_4\theta^{(i+2)}_4 &(i=1,2)&\longrightarrow &
\theta^{(i)}_2\theta^{(i+1)}_2\theta^{(i+2)}_4 -
\theta^{(i)}_4\theta^{(i+2)}_2\theta^{(i+3)}_2 & 2\ ^\nearrow & 2q^{-3} \\
\hline
\eta^{(i)}_2\theta^{(i+1)}_6 &(i=1,\ldots,4)&\longrightarrow &
\eta^{(i)}_2\Big(\theta^{(i+1)}_2\theta^{(i+2)}_4 + \theta^{(i+1)}_4\theta^{(i+3)}_2\Big) &
4 \longrightarrow 4 & \\
\theta^{(i)}_2\eta_6^{(i+1)} &(i=1,\ldots,4)&\longrightarrow &
\theta^{(i)}_2\Big(\eta^{(i+1)}_2\theta^{(i+2)}_4+\theta^{(i+1)}_2\eta^{(i+2)}_4 +
& \!\!\!4 \ \stackrel{^\nearrow}{\phantom{_\searrow}} & \\
&&& + \eta^{(i+1)}_4\theta^{(i+3)}_2+\theta^{(i+1)}_4\eta^{(i+3)}_2 \Big)&\ \searrow&4\\
\eta^{(i)}_4\theta^{(i+2)}_4 &(i=1,\ldots,4)&\longrightarrow &
\Big(\eta^{(i)}_2\theta^{(i+1)}_2+\theta^{(i)}_2\eta^{(i+1)}_2\Big)\theta^{(i+2)}_4 -
&4\longrightarrow 4 &\\
&&& -\eta^{(i)}_4\theta^{(i+2)}_2\theta^{(i+3)}_2 &&\\
\hline
\eta^{(i)}_2\eta^{(i+1)}_6 &(i=1,\ldots,4)&\longrightarrow &
\eta^{(i)}_2\Big(\eta^{(i+1)}_2\theta^{(i+2)}_4+\eta^{(i+1)}_4\theta^{(i+3)}_2
& 4 \longrightarrow 4 & 0 \\
&&& + \theta^{(i+1)}_2\eta^{(i+2)}_4 +\theta^{(i+1)}_4\eta^{(i+3)}_2\Big) &&\\
\eta^{(i)}_4\eta^{(i+2)}_4 &(i=1,2)&\longrightarrow &
\Big(\eta^{(i)}_2\theta^{(i+1)}_2+\theta^{(i)}_2\eta^{(i+1)}_2\Big)\eta^{(i+2)}_4
& 2 \longrightarrow 2 & 0\\
&&& - \eta^{(i)}_4\Big(\eta^{(i+2)}_2\theta^{(i+3)}_2+\theta^{(i+2)}_2\eta^{(i+3)}_2\Big)&&
\end{array}
\label{ta4d3}
\ee
This time the numbers in the last column can need a more detailed explanation.
To see what happens, we introduce a brief notation for four different structures,
arising in the image of $d_3$:
\be
\theta^{(i)}_4\theta^{(i+2)}_2\theta^{(i+3)}_2 = (422)_i = \alpha_i, \nn\\
\theta^{(i)}_4\eta^{(i+2)}_2\theta^{(i+3)}_2 = (4\bar 22)_i = \beta_i, \nn\\
\theta^{(i)}_4\theta^{(i+2)}_2\eta^{(i+3)}_2 = (42\bar 2)_i = \gamma_i, \nn\\
\eta^{(i)}_4\theta^{(i+2)}_2\theta^{(i+3)}_2 = (\bar 422)_i = \delta_i
\ee
(the image of the $\eta\eta$ sector is simple, and does not require additional
comments).
Then, for example,
\be
\begin{array}{ccc}
\theta^{(1)}_2\theta^{(2)}_6 &\longrightarrow & -\alpha_3-\alpha_2\\
\theta^{(1)}_4\theta^{(3)}_4 &\longrightarrow & \alpha_3-\alpha_1\\
\theta^{(1)}_6\theta^{(4)}_2 = -\theta^{(4)}_2\theta^{(1)}_2
&\longrightarrow & \alpha_6+\alpha_5 = \alpha_2+\alpha_1
\end{array}
\ee
The third line is the same as the first one, with $i=1$ shifted to $i+3$.
The sum of the three lines is the $d_3$-transform of the $d_2$-image of $\theta^{(1)}_8$,
thus it should vanish (in the last column) -- and, obviously, does so.
Transforms of the six individual terms in the sector $\theta\theta$
are all of the form $\alpha_j+\alpha_{j+1}$ or $\alpha_j-\alpha_{j+2}$, i.e. all
are linear combinations of the three independent structures, say,
$\alpha_1+\alpha_2$, $\alpha_2+\alpha_3$ and $\alpha_3+\alpha_4$.
This gives the final quantity $3q^{-2}$ in the last column of the table
(\ref{ta4d3}).

Similarly, in the $\theta\eta$ sector the structures, emerging in the image are:
$u_j=\beta_{j+1}+\gamma_{j}$, $v_j=\beta_j+\gamma_{j+1}+\delta_j+\delta_{j+1}$
and $w_j=\beta_j+\gamma_j-\delta_{j+2}$.
However, from these twelve quantities only eight are linear independent.
For example,
\be
d_3\Big(d_2(\eta^{(1)}_8)\Big) = d_3\Big(\eta^{(1)}_2\theta^{(2)}_6
+ \eta^{(1)}_4\theta^{(3)}_4 + \eta^{(1)}_6\theta^{(4)}_4 +
\theta^{(1)}_2\eta^{(2)}_6
+ \theta^{(1)}_4\eta^{(3)}_4 + \theta^{(1)}_6\eta^{(4)}_4\Big) =
\ee
\vspace{-0.4cm}
$$
= (-\beta_3-\gamma_2) + (\beta_3+\gamma_3-\delta_1) + (\beta_1+\gamma_2+\delta_1+\delta_2)
- (\beta_2+\gamma_3+\delta_2+\delta_3) - (\beta_1+\gamma_1-\delta_3) + (\beta_2+\gamma_1)
= 0
$$
This seems to provide four elements of ${\rm Ker}(d_3)$, but in fact there is one
linear relation between the four, so actually only three remain -- this was the
reason why $3$ appears as the $\theta\eta$-sector contribution to ${\rm dim}_q{\rm Ker}(d_2)$.
However, $d_3$ has one extra zero mode. Indeed, all $v$-variables are expressed through
$u$ and $w$:
\be
v_j = w_{j+1}+w_j-u_j +\sum_{k=1}^4(u_k-w_k)
\ee
Thus
\be
n=4:\ \ \ \ {\rm dim}_q{\rm Ker}(d_3) = 3q^{-2} + 4
\ \stackrel{(\ref{dimimd2})}{=}\  {\rm dim}_q{\rm Im}(d_2)
\ \ \Longrightarrow \ \ \ {\rm dim}_q(H_2^{(4)}) = 1
\ee
and the $T^2$ contribution to $P^{(4)}(T,q)$ is $q^4(qT)^2\cdot 1 = q^6T^2$.

The answer remains structurally the same for all other $n\geq 3$:
\be
{\rm dim}_q{\rm Ker}(d_3) = (n-1)q^{-2} + n \ \
\Longrightarrow \ \ {\rm dim}_q(H_2^{(n)}) = 1
\ee
the contribution to $P^{(n)}(T,q)$ is $q^n(qT)^2 = q^{n+2}T^2$
and
\be
{\rm dim}_q{\rm Im}(d_3) = q^{-1}\Big({\rm dim}_q(C_2) -{\rm dim}_q{\rm Ker}(d_3)\Big)
= q^{-1}\left(\frac{n(n-1)}{2}D^2 - (n-1)q^{-2} - n\right)
= \nn \\
= \frac{(n-1)(n-2)}{2}\,q^{-3} + n(n-2)q^{-1} + \frac{n(n-1)}{2}\,q
\ee

\subsubsection{Differential $d_I$ with generic $I$}

Once (\ref{d2brab}) and (\ref{d3brab}) are known, it is straightforward
to write down the full BRST operator for the case of arbitrary $2$-strand
braid with all black vertices:
\be
\boxed{
\  {\cal Q}^{(n)} = \sum_{I=1}^n d_I^{\,(n)} =
\left(\sum_{i=1}^n \vartheta_{2n}^{(i)}\right)\frac{\p^2}{\p\vt_n\p\vt_n'}
-\sum_{i,j,k}^n \vt^{(i)}_{2j-2i}\vt^{(j)}_{2k-2j}\frac{\p}{\p\vt^{(i)}_{2k-2i}}\
}
\label{Qn}
\ee
The first term is the $d_1$ from (\ref{d1bbbb}), and the second term depends on $n$
also through the agreement $\vt^{(i)}_{-2l} = \vt^{(i)}_{2n-2l}$.
Particular differential $d_I$ arises automatically when this operator is restricted
to the relevant space $C_{I-1}$.
Conditions $d_{I+1}d_I = 0$ follow from the nilpotency of ${\cal Q}^{(n)}$,
$\Big({\cal Q}^{(n)}\Big)^2=0$,
which, in turn, depends on the properties (\ref{co}) and (\ref{coco}) of
the tensors $Q_{ij}^k$ and $Q^{ij}_k$, implicitly present in (\ref{Qn})
through the convention (\ref{elopsusy}).
Given these properties, we can write:
\be
\Big({\cal Q}^{(n)}\Big)^2=
\sum_{i,j,k}
\vt^{(i)}_{2j-2i}\vt^{(j)}_{2k-2j}\frac{\p}{\p\vt^{(i)}_{2k-2i}}\cdot
\sum_{i',j',k'}
\vt^{(i')}_{2j'-2i'}\vt^{(j')}_{2k'-2j'}\frac{\p}{\p\vt^{(i')}_{2k'-2i'}} = \nn \\
= \sum_{\stackrel{i,j,k}{i',j',k'}} \left(\delta{^{i'}_i}\delta_{j'-i'}^{k-i}\cdot
\vt^{(i)}_{2j-2i}\vt^{(j)}_{2k-2j}\vt^{(j')}_{2k'-2j'}\frac{\p}{\p\vt^{(i')}_{2k'-2i'}}
\ -\ \delta{^{j'}_i}\delta_{k'-j'}^{k-i}\cdot
\vt^{(i)}_{2j-2i}\vt^{(j)}_{2k-2j} \vt^{(i')}_{2j'-2i'}\frac{\p}{\p\vt^{(i')}_{2k'-2i'}}
\right) = \nn \\ =
\sum_{i,j,k,k'}
\vt^{(i)}_{2j-2i}\vt^{(j)}_{2k-2j}\vt^{(k)}_{2k'-2k}\frac{\p}{\p \vt^{(i)}_{2k'-2i}}
\ -\sum_{i,j,k,i'}
\vt^{(i)}_{2j-2i}\vt^{(j)}_{2k-2j} \vt^{(i')}_{2i-2i'}\frac{\p}{\p\vt^{(i')}_{2k-2i'}}
= 0
\ee
where at the last stage we make a change of summation variables in the second sum:
$i' = i,\ k=k',\ i=j,\ j=k$, which converts it into the first one.

For evaluation of cohomologies convenient is the second,
more explicit, version of formula (\ref{d3brab}).

\subsubsection{Differential $d_n$}

$d_n$ acts into $2^n$-dimensional space with basis $\otimes_{i=1}^n \vt_2^{(i)}$
from $2^{m-1}n$-dimensional one with the basis
$\oplus_{i=1}^n \vt_4^{(i)}\otimes_{j\neq i,i+1}\vt_2^{(j)}$:
\be
d_n = \sum_{i=1}^n \vt_2^{(i)}\vt_2^{(i+1)}\frac{\p}{\p\vt_4^{(i)}}
= \sum_{i=1}^n \theta_2^{(i)}\theta_2^{(i+1)} \frac{\p}{\p\theta_4^{(i)}}+
\sum_{i=1}^n
(\theta_2^{(i)}\eta_2^{(i+1)}+\eta_2^{(i)}\theta_2^{(i+1)}) \frac{\p}{\p\eta_4^{(i)}}
\ee
Clearly, coimage of $d_n$ contains $\prod_{i=1}^n \eta_2^{(i)}$
which has $q$-grading $q^n$.
For $n$ odd (knots) this is the only term in $\ {\rm Coim}(d_n)$,
however, for even $n$ (links) there is one more:
$\sum_{i=1}^n (-)^i\theta_2^{(i)}\prod_{j\neq i}\eta_2^{(j)}$.
This means that the $T^n$-term in the superpolynomial is equal to
\be
\left\{
\begin{array}{rcc}
0 & {\rm for} & n=1 \\
q^n(qT)^nq^n=q^{3n}T^n &\ {\rm for\ odd} \ &n>1\\
q^n(qT)^n(q^{n-2}+q^n) = q^{3n-2}T^n + q^{3n}T^n = q^{3n-1}DT^n&\ {\rm for\ even}\ &n
\end{array}\right.
\ee

As to the $T^{n-1}$ term, it is controlled by the kernel of $d_n$.
In particular no $\eta^{n-1}$ can appear in the image of $d_{n-1}$,
while the action of $d_n$ converts
\be
d_n: \ \ \ \eta^{(i)}_4\prod_{k=2}^{n-1} \eta^{(i+k)}_2 \ \longrightarrow
\ \ \
\Big(\theta_2^{(i)}\eta^{(i+1)}_2 + \eta_2^{(i)}\theta^{(i+1)}_2\Big)
\prod_{k=2}^{n-1} \eta^{(i+k)}_2\ =\ \alpha_i-(-)^{n}\alpha_{i+1}
\ee
where $\alpha_i = \theta^{(i)}_2\prod_{i=k}^{n-1} \eta_2^{(i+k)}$.
For even $n$ variables there is one linear relation between the structures
at the r.h.s.
\be
\sum_{i=1}^n (\alpha_i - \alpha_{i+1}) = 0
\ee
for odd $n$ they are all linearly independent.
This implies that the contribution of the order $q^n(qT)^{n-1}$
to the superpolynomial contains a $q$-grading factor $q^{n-1}$
for even $n$, but not for odd $n$.
In fact this means that for odd $n$ the single contribution to
$H_{n-1}$ comes from the sector $\theta\eta^{n-2}$, and this factor is
 rather $q^{n-3}$.

\subsubsection{Full answer}

It turns out that in our example of $2$-strand braids will all black vertices
cohomologies $H_k^{(n)}$ with $1<k<n$
are always 1-dimensional, and lie either in one of the two sectors:
either $\eta^{k-1}\theta$ or $\eta^k$.
Actually,
\be
{\rm dim}_q\Big(H_k^{(n)}\Big) = \left\{
\begin{array}{cc}
q^{k-2} & {\rm for\ even}\ k \\
q^k & {\rm for\ odd}\ k
\end{array}\right.
\ee
Putting all pieces together, we finally obtain for
the unreduced Jones superpolynomial  in the fundamental representation
for the torus knot/link of the type $[2,n]$:
$$
P^{(n)}_{_\Box}(T,q) \stackrel{\cite{Khof,superpolsfirst}}{=}
q^n \left( q^{-1}DT^0 + 0\cdot(qT)^1 + \sum_{{\rm even}\ k=2}^{n-1}
q^{k-2}(qT)^k  + \sum_{{\rm odd}\ k=3}^{n-1} q^k(qT)^k
+ \left(\begin{array}{c}q^{n} \\ q^{n-1}D\end{array}\right) (qT)^n
\right) =
$$
\vspace{-0.5cm}
\be
=\ q^n  \left( q^{-2}+1+q^2T^2 + \
(1+T)\!\!\!\! \sum_{{\rm odd}\ k=3}^{n-1}\!  q^{2k}T^k
+ q^{2n}T^n\right)
\label{ansJ}
\ee
For $T=-1$ this turns back into the $4$-term fundamental Jones
(\ref{Jn})
However, starting from $n=4$  the unreduced Jones superpolynomial (\ref{ansJ})
acquires {\it more} terms than original Jones (more than four), there is a non-trivial
contribution, proportional to $(T+1)$: this is a signal that new
topological information occurs after $T$-deformation,
not seen at all at $T\neq -1$.
(Actually, the possibility to see this phenomenon at the level of $2$-strand
braids is the specifics of {\it non-reduced} knot polynomials.
To observe it in the case of {\it reduced} polynomials,
one should go beyond $2$-strands.)

\bigskip

We can compare (\ref{ansJ}) with concrete answers, listed in \cite{CM}.
They are given there for arbitrary $N$, we need $N=2$ --
this is the third column in the following table
(we also omit lengthy expression for $n=4$, while for $n=6$
ref.\cite{CM} in any case gives it only for $N=2,3$;
note also that torus links with $n=4\ \&\ 6$ are labeled "v2" in \cite{CM}
-- "v1" differ by orientation of one of the two components of the link
and are not torus):
\be
\begin{array}{cccc}
n & P^{(n)}(T,q|N) & P^{(n)}(T,q) & {\rm ref} \\ \\
\hline \\
2: & q^{N-1}[N] + (q^{2N}[N]-q^{N+1})[N]T^2 & qD(1+q^4T^2) & \cite{GIKV}   \\
3:& q^{2(N-1)}\Big([N]+[N-1]q^3T^2(1+q^{2N}T)\Big) & q+q^3+q^5T^2+q^9T^3 & \cite{RJ} \\
4: &   & q^2+q^4+q^6T^2 + q^{10}T^3 + q^{11}DT^4
&  \cite{CM}\\
5: & q^{4(N-1)}\Big([N]+[N-1]q^3T^2(1+q^{2N}T)(1+q^4T^2)\Big)&
q^3+q^5 +q^7T^2 + q^{11}T^3+q^{11}T^4+q^{15}T^5
& \cite{RJ}  \\
6: &
& q^4+q^6+ q^{8}T^2+q^{12}T^3+q^{12}T^4+q^{16}T^5 + q^{17}DT^6
&   \cite{CM}\\
\ldots
\end{array}
\nn
\ee

\subsection{A technical comment}

Note that in all above examples Grassmannian nature of variables
did not play too much {\it technical} role: it affects only signs,
while nilpotency is not really important -- it is automatically
taken into account by the choice of the vector spaces to act from
and to. This means that in cohomology calculations one can
actually work directly with the cut-and-join operator (\ref{Wop}),
what is considerably easier in more complicated examples.

Also, in Appendix B at the end of this paper we give a more traditional
description of the simplest examples, representing the differentials
as explicit rectangular matrices -- this, however, is a difficult
approach if one needs to consider {\it series} of knots/links,
say, all the $2$-strand braids at once (what is needed, for example,
to compare with the {\it evolution} method of \cite{DMMSS}).

\section{Conclusion}

In this introductory review of Khovanov's approach to the superpolynomials
we discussed the basic example of the unreduced Jones superpolynomial
and described in some detail the simplest case of the $2$-strand knots.
However, we did not yet manage to find an acceptably clear explanation
(desirably free of the homological algebra techniques)
of the topological invariance, which should separate the locality principle
and local restrictions, imposed on differential operators by invariance under
the \Red moves.

The $2$-strand example is sufficient to demonstrate the usual
peculiarity of the homological approach:
explicit construction of all the relevant spaces and operators
is very tedious -- but at the same time transparent and absolutely algorithmic.
This makes it easy to get {\it answers} with the help of computer simulations
(though computer abilities are exponentially decreasing with the growth
of the crossing numbers),
but pretty difficult to study any theoretical issues, like searching
for hidden {\it structures}, getting one-parametric {\it families} of formulas etc
-- it is difficult to keep any parameters unfixed.

Construction includes the substitution of original {\bf link diagram}
(a $4$-valent oriented graph with colored vertices of two types)
by a {\bf hypercube} of its resolutions, which is then promoted to a
hypercube {\bf quiver} and projected into a {\bf complex} with differentials,
made from the quiver maps.
The superpolynomial is the {\bf Poincare polynomial} if the complex.
To construct the differentials in the complex we used the
Koszul counterpart of the {\bf cut-and-join operator} and
{\bf extended Jones polynomial}:
looking at the latter one can pick up the appropriate items of the former.
The main freedom in the construction is the quiver structure:
the choice of the vector spaces at the hypercube vertices
(associated with connected components of the graph resolutions)
and the choice of the linear maps between them.
We did not describe explicitly the restrictions, imposed on these
spaces and maps by locality principle (none?)
by appearance of the complex  in the projection
(commutativity and associativity)
and by the \Red invariance.

In the following parts of this review series we are going to
make this part of the story more explicit and
address a number of obvious issues:

(i) Generalization to

$\bullet$ arbitrary groups $Sl(N)$

$\bullet$ arbitrary representations

(ii) Lifting to universal polynomials

$\bullet$ $N \longrightarrow A=q^N$

$\bullet$ $T \longrightarrow t=-qT = q^\beta$

(iii) Generalization to

$\bullet$ {\it reduced} superpolynomials, admitting character decomposition
in MacDonald dimensions \cite{DMMSS},

$\bullet$ their further lifting to {\it extended} superpolynomials,
where dimensions are substituted by full MacDonald polynomials,
depending on infinitely many "time-variables" $\{p_k\}$ \cite{DMMSS,AMMkn1}.
(these are invariants of braids only, not of knots/links -- but this
is already the case for some versions of Khovanov-Rozansky construction
for $N>2$).

(iv) Connection to index theorems and their reformulations
in terms of
SUSY quantum mechanics {\it a la} \cite{NP}.

(v) Relations with other approaches to superpolynomial calculus.

\section*{Appendix A: Sketch of  invariance proof at the
hypercube level}

Each reformulation of Jones polynomial formula in sections 1-5
represents it in a different form, and, more important,
in different terms.
Therefore the proof or \Red invariances at each level
looks different and deserves separate discussion.
Of course, existence of the most transparent proof
in algebraic terms (sec.1) is sufficient, but,
unfortunately, such formulation is not yet available
for superpolynomials.
On the other hand,
relevant for superpolynomial deformation
is reformulation in quiver terms (secs.3-4),
which we did not yet manage to present in adequate terms
(for numerous homological-algebra presentations of
the proof see \cite{Khof}--\cite{Khol}).
Therefore we put accents away from invariance proofs
in the main text.
Still, in this appendix we describe a proof in terms
of the intermediate stage: at the very important level
of the hypercube of graph resolutions (sec.2).
It helps to better understand, how the notion of {\it locality}
-- trivial at the algebraic level of sec.1 --
gets non-trivial in all other formulations.
Still at this level it is easy to present non-formally,
what makes this intermediate example instructive and
potentially useful.

\subsection*{$R1$ move}

\noindent

$\bullet$ Take a graph $L$ with $N$ vertices and a polynomial $J_{L}(q)$.

$\bullet$ Cut an edge and insert a "loop", with either a white or a black vertex:

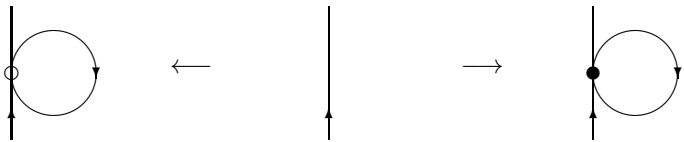
\begin{picture}(300,100)(-150,-50)
\put(0,-25){\line(0,1){50}}
\put(0,-15){\vector(0,1){2}}
\put(100,-25){\line(0,1){50}}
\put(116,0){\circle{30}}
\put(100,0){\circle*{5}}
\put(100,-15){\vector(0,1){2}}
\put(132,0){\vector(0,-1){2}}
\put(-120,-25){\line(0,1){50}}
\put(-104,0){\circle{30}}
\put(-120,0){\circle{5}}
\put(-120,-15){\vector(0,1){2}}
\put(-88,0){\vector(0,-1){2}}
\put(50,0){$ {\longrightarrow}$}
\put(-60,0){$ {\longleftarrow}$}
\end{picture}

$\bullet$ For the graph $\Gamma$ this means just insertion of a new uncolored vertex:

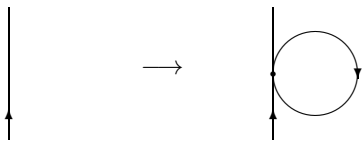
\begin{picture}(300,100)(-150,-50)
\put(0,-25){\line(0,1){50}}
\put(0,-15){\vector(0,1){2}}
\put(100,-25){\line(0,1){50}}
\put(116,0){\circle{30}}
\put(100,0){\circle*{2}}
\put(100,-15){\vector(0,1){2}}
\put(132,0){\vector(0,-1){2}}
\put(50,0){$ {\longrightarrow}$}
\end{picture}

$\bullet$ Adding a vertex to a graph $\Gamma$ means doubling the hypercube $\Box(\Gamma)$.

The first copy contains a new resolved vertex of the type $\ =\ $:

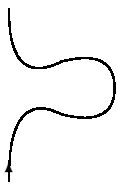
\begin{picture}(300,100)(-150,-50)
\qbezier(0,-35)(0,0)(20,-10)
\qbezier(20,-10)(40,-15)(40,0)
\qbezier(20,10)(40,15)(40,0)
\qbezier(0,30)(0,0)(20,10)
\put(0,-30){\vector(0,1){2}}
\end{picture}

The second copy -- of the type $\ ||\ $:

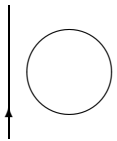
\begin{picture}(300,100)(-150,-50)
\put(0,-25){\line(0,1){50}}
\put(0,-15){\vector(0,1){2}}
\put(23,0){\circle{30}}
\end{picture}

$\bullet$ The first insertion does not change $J_{L}(q)$,
the second adds one new disconnected component to the graph,
and multiplies $J_{L}(q)$ by $\gamma$.

$\bullet$ Another thing that changes is the initial vertex.

If we insert white vertex, then the first copy of the cube
goes first -- with the coefficient one, -- while the second
copy goes second -- with the coefficient $-q$.
And a new factor $\alpha$ is added.

If we insert the black vertex, then first -- with the coefficient one
-- goes the second copy of the cube, while the first copy
of the cube goes second -- with the coefficient $-q$.
And added is a factor $\beta$.

$\bullet$ Thus insertion of a loop with a white vertex substitutes
the polynomial
\be
J_{L}(q) \ \longrightarrow \ \alpha(1-q\gamma)J_{L}(q)
\ee
while insertion of a loop with a black vertex --
\be
J_{L}(q) \ \longrightarrow \ \beta(\gamma-q)J_{L}(q)
\ee

$\bullet$ Invariance under the \Red move $R1$ requires that these two factors
are unity:
\be
\alpha(1-q\gamma) = 1, \nn \\
\beta(\gamma-q) = 1
\ee
what expresses $\alpha$ and $\beta$ through $\gamma$ and $q$.

\subsection*{$R2$ move}

\noindent

$\bullet$ Take a graph $L$ with $N$ vertices and a polynomial $J_{L}(q)$.

$\bullet$ Cut two edges. This provides a new graph $L_{ab|cd}$
with four external legs, two incoming ($a,b$) and two outgoing ($c,d$).
Connecting $c$ with $a$ and $d$ with $b$ gives original graph
$L_{ab|ab} = L$. Two other connections give new graphs:
non-oriented $M=L_{aa|cc}$ and oriented $N=L_{ab|ba}$.
The polynomials $J_M(q)$ and $J_N(q)$ in general are different from $J_L(q)$.

\bigskip

Now we can go in different directions, by making different kind of insertions
between external legs.
We begin with the Redemeister move $R2$, and continue with the skein
(Hecke algebra) relation in the next subsection.

\bigskip

$\bullet$ Insert a pair of a white and a black vertices:

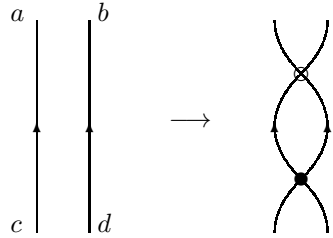
\begin{picture}(300,100)(-150,-50)
\put(0,-40){\line(0,1){80}}
\put(20,-40){\line(0,1){80}}
\put(-10,40){\mbox{$a$}}
\put(-10,-40){\mbox{$c$}}
\put(23,40){\mbox{$b$}}
\put(23,-40){\mbox{$d$}}
\put(0,0){\vector(0,1){2}}
\put(20,0){\vector(0,1){2}}
\put(50,0){\mbox{$\longrightarrow$}}
\put(100,20){\circle{5}}
\put(100,-20){\circle*{5}}
\qbezier(100,20)(90,30)(90,40)
\qbezier(100,20)(110,30)(110,40)
\qbezier(100,-20)(90,-30)(90,-40)
\qbezier(100,-20)(110,-30)(110,-40)
\qbezier(100,20)(80,0)(100,-20)
\qbezier(100,20)(120,0)(100,-20)
\put(90,0){\vector(0,1){2}}
\put(110,0){\vector(0,1){2}}
\end{picture}

$\bullet$
There are four different resolutions:

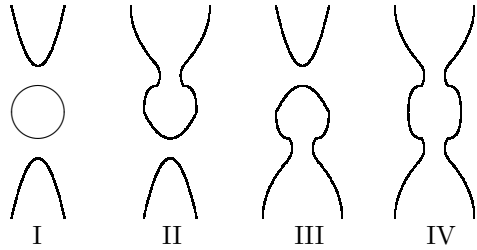
\begin{picture}(300,110)(-150,-55)
\qbezier(-50,40)(-40,-5)(-30,40)
\put(-40,0){\circle{20}}
\qbezier(-50,-40)(-40,5)(-30,-40)
\put(-42,-50){\mbox{I}}
\qbezier(-5,40)(-5,30)(3,20)
\qbezier(3,20)(8,15)(5,10)
\qbezier(5,10)(0,10)(0,0)
\qbezier(0,0)(10,-20)(20,0)
\qbezier(25,40)(25,30)(17,20)
\qbezier(17,20)(12,15)(15,10)
\qbezier(15,10)(20,10)(20,0)
\qbezier(0,-40)(10,5)(20,-40)
\put(7,-50){\mbox{II}}
\put(57,-50){\mbox{III}}
\put(107,-50){\mbox{IV}}
\qbezier(50,40)(60,-5)(70,40)
\qbezier(45,-40)(45,-30)(53,-20)
\qbezier(53,-20)(58,-15)(55,-10)
\qbezier(55,-10)(50,-10)(50,0)
\qbezier(50,0)(60,20)(70,0)
\qbezier(75,-40)(75,-30)(67,-20)
\qbezier(67,-20)(62,-15)(65,-10)
\qbezier(65,-10)(70,-10)(70,0)
\qbezier(95,40)(95,30)(103,20)
\qbezier(103,20)(108,15)(105,10)
\qbezier(105,10)(100,10)(100,0)
\qbezier(125,40)(125,30)(117,20)
\qbezier(117,20)(112,15)(115,10)
\qbezier(115,10)(120,10)(120,0)
\qbezier(95,-40)(95,-30)(103,-20)
\qbezier(103,-20)(108,-15)(105,-10)
\qbezier(105,-10)(100,-10)(100,0)
\qbezier(125,-40)(125,-30)(117,-20)
\qbezier(117,-20)(112,-15)(115,-10)
\qbezier(115,-10)(120,-10)(120,0)
\end{picture}

$\bullet$
Clearly, the first three resolutions
will substitute $J_L(q)$ by $J_M(q)$,
while the forth resolution will leave
$J_L(q)$ intact.
Also the first resolution adds
one new disconnected component and thus a factor $\gamma$.

$\bullet$
Insertion of
two white vertices means that the resolution $I$ goes first
(i.e. with coefficient one),
the two resolutions $II$ and $III$ go second (i.e. with the coefficient $-q$),
the resolution $IV$ goes third (with the coefficient $q^2$).
This means that
\be
J_L(q) \ \longrightarrow \
\alpha^2\Big( \underbrace{\gamma J_M(q)}_I
-q\cdot\underbrace{2J_M(q)}_{II+III} +q^2 \underbrace{ J_L(q)}_{IV}\Big)
\ee
Similarly, insertion of two black vertices would give
\be
J_L(q) \ \longrightarrow \
\beta^2\Big( \underbrace{ J_L(q)}_{IV}
-q\cdot\underbrace{2J_M(q)}_{II+III} + q^2\underbrace{\gamma J_M(q)}_{I}\Big)
\ee

$\bullet$ Insertion one white and one black vertex at the
first and second place respectively implies that the first
-- with the coefficient one -- goes the resolution $II$,
the second -- with the coefficient $(-q)$ -- go $I$ and $IV$,
and the last -- with the coefficient $q^2$ -- goes $III$, so that
\be
J_L(q) \ \longrightarrow \
\alpha\beta\left(
\underbrace{J_M(q)}_{II} - q\cdot\Big(\underbrace{\gamma J_M(q)}_I +
\underbrace{J_L(q)}_{IV}\Big) + q^2\underbrace{J_M(q)}_{III}
\right) = -q\alpha\beta J_L(q) +
\alpha\beta\Big(1-q\gamma+q^2\Big)J_M(q)
\ee
Similarly, if the black vertex goes first and the white goes second, we have:
\be
J_L(q) \ \longrightarrow \
\alpha\beta\left(
\underbrace{J_M(q)}_{III} - q\cdot\Big(\underbrace{\gamma J_M(q)}_I +
\underbrace{J_L(q)}_{IV}\Big) + q^2\underbrace{J_M(q)}_{II}
\right) = -q\alpha\beta J_L(q) +
\alpha\beta\Big(1-q\gamma+q^2\Big)J_M(q)
\ee

$\bullet$
Invariance under the $R2$ \Red move implies that
$J_L(q)$ remains intact when the pair of black and white vertices
is inserted, this means that the coefficient in front of $J_L(q)$
should be one, and that in front of $J_M(q)$ vanishes:
\be
-q\alpha\beta  = 1, \nn \\
1-q\gamma+q^2 = 0
\ee

\subsection*{Skein relation}

\noindent

$\bullet$
Going back, to the very beginning of the previous subsection,
instead of a pair of vertices we could insert just one, and check the skein
(or Hecke-algebra) relation
\be
q^2\circ - q^{-2}\,\bullet\ =\ q-q^{-1}
\ee
(this is  the property of the ${\cal R}$-matrix in the fundamental representation).

$\bullet$ When white vertex is inserted, we get:
\be
J_L(q) \ \longrightarrow \ \alpha\Big(J_M(q) - qJ_L(q)\Big)
\ee
When black vertex is inserted, we get instead
\be
J_L(q) \ \longrightarrow \ \beta\Big(J_L(q) - qJ_M(q)\Big)
\ee

$\bullet$ Thus skein relation implies that
\be
q^2 \alpha\Big(J_M(q) - qJ_L(q)\Big) - q^{-2} \beta\Big(J_L(q) - qJ_M(q)\Big)
= (q-q^{-1})J_L(q)
\ee
i.e.
\be
\alpha q^2 + \beta q^{-1} = 0, \nn \\
-\alpha q^3 - \beta q^{-2} = q-q^{-1}
\ee

\subsection*{$R3$ move}

\noindent

$\bullet$ Take a graph $L$ with $N$ vertices and a polynomial $J_{L}(q)$.

$\bullet$ Cut three edges, obtain the graph $L_{abcdef}$ with six ends,
and insert a triple intersection in one of two ways:

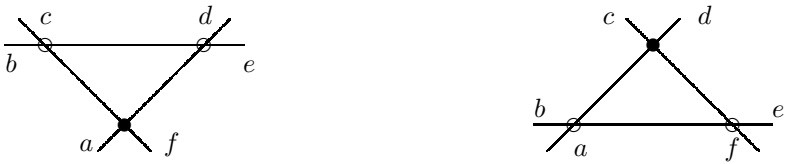
\begin{picture}(300,90)(-50,-30)
\qbezier(-10,-10)(10,10)(40,40)
\qbezier(10,-10)(-10,10)(-40,40)
\qbezier(-45,30)(0,30)(45,30)
\put(0,0){\circle*{5}}
\put(-30,30){\circle{5}}
\put(30,30){\circle{5}}
\qbezier(190,40)(210,20)(240,-10)
\qbezier(210,40)(190,20)(160,-10)
\qbezier(155,0)(200,0)(245,0)
\put(170,0){\circle{5}}
\put(230,0){\circle{5}}
\put(200,30){\circle*{5}}
\put(-17,-10){\mbox{$a$}}
\put(-45,20){\mbox{$b$}}
\put(-32,38){\mbox{$c$}}
\put(28,38){\mbox{$d$}}
\put(45,20){\mbox{$e$}}
\put(15,-10){\mbox{$f$}}
\put(170,-12){\mbox{$a$}}
\put(155,3){\mbox{$b$}}
\put(181,38){\mbox{$c$}}
\put(217,38){\mbox{$d$}}
\put(245,3){\mbox{$e$}}
\put(227,-12){\mbox{$f$}}
\end{picture}

\noindent
In fact, the black vertex could also be white -- what matters
is that the two vertices on the horizontal line are of the same color.

$\bullet$ Each of the two pictures will have eight resolutions.

$\bullet$
When attached to $L_{abcdef}$ these sixteen resolutions will
give rise to just five new graphs,
\be
A = L_{abbdda},\ \ \ B=L_{abbaee},\ \ \ C=L_{abccba},
\ \ \ D = L_{aacddc},\ \ \ E=L_{aaccee}
\ee
Equality between the two pictures
(invariance under the $R3$ move, i.e. the Yang-Baxter relation)
requires that coefficients coincide in front of $A,B,C,D,E$
for both pictures. For $B,C,D$ this happens automatically,
and non-trivial are the conditions for $A$ and $E$ only,
this gives two constraints:
\be
1-q\gamma+q^2=0, \nn \\
q^2\gamma - q - q^3 = 0
\ee
both satisfied for $\gamma = q+q^{-1}$.

$\bullet$ It is important here that there no graph
$L_{abcabc}$ appears in this process,
this would provide one more constraint which would be
impossible to satisfy.
This does not allow to add a third type of resolution
(crossing) into this whole construction.

\subsection*{Example of topological invariance: changing black to white
in the $2$-strand braid}

If some $m$ out of $n$ black vertices in the $2$-strand braid are changed for white,
this provides a link, topologically equivalent to the one with $n-m$ black vertices
only.

$\underline{n_\circ = 1,\  n_\bullet=n-1:}\ $
Let us begin with the case of $m=1$ and let us change the color of the vertex number $I$.
Then this means that the starting vertex of the hypercube is now $[\underline{1}000\ldots0]$
instead of $[0000\ldots0]$. For convenience we underline this distinguished $1$.
The first-level vertices, instead of $[0\ldots 010\ldots 0]$
with a single unity at some place, are now substituted by
$[\underline{0}00\ldots 0]$ and $[\underline{1}0\ldots 010\ldots 0]$.
Similarly, at level $J$, instead of the $\frac{n!}{J!(n-J)!}$ hypercube vertices
with $J$ unities, we now get
$\frac{(n-1)!}{J!(n-1-J)!}$ vertices with unity at the first position and
$J$ unities somewhere else,
as well as $\frac{(n-1)!}{(J-1)!(n-J)!}$ vertices with $0$ at the first position
and $J-1$ unities somewhere else.
Since in the case of $2$-strand braids the number of connected components of
resolved graph is equal to the number of unities in the label of hypercube vertex
-- with the single exception of $[00\ldots 0]$ for which this number is two,
we get:
\be
J^{n_\circ=1,\,n_\bullet=n-1}_{_\Box}(q)
= -q^{-2}\cdot q^{n-1}\Big\{\underline{D} - q\Big(\underline{(n-1)D^2}+\boxed{D^2}\,\Big)
+ q^2\left(\frac{1}{2}(\underline{n-1)(n-2) D^3}+(n-1)D\right) - \nn \\
- q^3\left(\frac{1}{6}(\underline{n-1)(n-2)(n-3)D^4} + \frac{1}{2}(n-1)(n-2)D^2\right)
+ \ldots \Big\} = \nn \\
= -q^{n-3}\Big\{\underline{D(1-qD)^{n-1}} +\boxed{q(1-D^2)} -q(1-qD)^{n-1}\Big\} = \nn \\
= -q^{n-3}\Big\{(D-q)(1-qD)^{n-1} - q(D^2-1)\Big\}
= q^{n-2}\Big(q^2+1+q^{-2} +(-q^2)^{n-2}\Big)
\stackrel{(\ref{Jn})}{=}\  J^{(n-2)}_{_\Box}(q)
\label{J1n-1}
\ee
what is the right answer for Jones for the $2$-strand braid with $n_\bullet-n_\circ =n-2$
uncompensated crossings. The answer is composed from two series of items (one of them
is underlined for convenience) and a "defect" (boxed), associated with "anomalous"
hypercube vertex $[00\ldots 0]$: it would smoothly get into the non-underlined series,
if contributed $D^0=1$, but actually it contributes $D^2$, and this defect should
be explicitly taken into account.

$\underline{n_\circ = 2,\  n_\bullet=n-2:}\ $
This time the starting vertex in the hypercube contains two units, let it be
$[\underline{11}00\ldots 0]$.
At the first level we have:
$[\underline{10}00\ldots  0]$, $[\underline{01}00 \ldots 0]$
and $[\underline{11}00\ldots 010\ldots 0]$.
At the second level appears the "anomalous" $\boxed{[\underline{00}00\ldots 0]}$
as well as $[\underline{10}00\ldots 010 \ldots 0]$, $[\underline{01}00\ldots 010 \ldots 0]$
and $[\underline{11}00\ldots 010 \ldots 010 \ldots  0]$.
At the third level we get
$[\underline{00}00\ldots 010 \ldots 0]$
$[\underline{10}00\ldots 010\ldots 010 \ldots 0]$,
$[\underline{01}00\ldots 010 \ldots 010 \ldots 0]$
and $[\underline{11}00\ldots 010 \ldots 010 \ldots 010 \ldots  0]$
and so on.
In result
\be
J^{n_\circ=2,\,n_\bullet=n-2}_{_\Box}(q)
= q^{-4}\cdot q^{n-2}\Big\{\underline{D^2} - q\Big(\underline{\underline{2D}}
+ \underline{(n-2)D^3}\Big) +
q^2\left(\boxed{D^2} + 2(n-2)D^2 + \frac{1}{2}(\underline{n-2)(n-3)D^4}\right) - \nn \\
- q^3\left((n-2)D + 2\cdot\frac{1}{2}(\underline{\underline{n-2)(n-3)D^3}}
+ \frac{1}{6}(\underline{n-2)(n-3)(n-4)D^5}\right) + \ldots
\Big\} = \nn \\
= q^{n-6}\Big\{ \underline{D^2(1-qD)^{n-2}} + \boxed{q^2(D^2-1)}
-\underline{\underline{2qD(1-qD)^{n-2}}} + q^2(1-qD)^{n-2}\Big\} = \nn \\
= q^{n-6}\Big\{q^2(D^2-1) +(D-q)^2(1-qD)^{n-2}\Big\}
= q^{n-4}\left(q^2+1+q^{-2} + (-q^2)^{n-4}\right)\
\stackrel{(\ref{Jn})}{=}\  J^{(n-4)}_{_\Box}(q)
\ee
as it should be.
Clearly, this time there are three series
(two of them underlined once and twice respectively) and one defect term.

Now it is clear that for {\it arbitrary $n_\circ$} we obtain $n_\circ + 1$ series
and the answer, implied by the hypercube pattern for the unreduced Jones polynomial, is:
\be
J^{n_\bullet=n-n_\circ}_{_\Box}(q) =
(-)^{n_\circ}q^{n_\bullet-2n_\circ}\Big\{
(-q)^{n_\circ}(D^2-1) + (D-q)^{n_\circ}(1-qD)^{n_\bullet} \Big\} = \nn \\
= q^{n_\bullet-n_\circ}\left(q^2+1+q^{-2} + (-q^2)^{n_\bullet-n_\circ}\right)\
\stackrel{(\ref{Jn})}{=}\  J^{(n_\bullet-n_\circ)}_{_\Box}(q)
\ee

\section*{Appendix B: Examples of cohomology calculus at the level of matrices}

We consider here the $2$-strand knots with one, two and three intersections
even more explicitly than it was done in ss.6 and 7.

\subsection*{Example: An eight}

Let us take just one black vertex. Then hypercube is just a segment
with two vertices $[0]$ and $[1]$, and
\be
C_0 = V\otimes V, \ \ \ C_1 = V
\ee
so that $J(q=1) = 2\cdot 2 - 2 =2$.
The differential
\be
d_1 = Q^* = \left(\begin{array}{cccc}
1 & 0 & 0 & 0 \\ 0 & 1 & 1 & 0
\end{array}\right)
\ee
It has rank two, therefore
\be
P(T,q)|_{q=1} = (4-2)T^0 + (2-2)T^1 = 2
\ee
The kernel of $d_1$ consists of two vectors $(01,-1,0)=v_+\otimes v_- - v_-\otimes v_+$
and $(0001) = v_-\otimes v_-$ with $q$-gradings $1$ and $q^{-2}$ respectively.
The coimage of $d_1$ is empty. Therefore
\be
P^\bullet(T,q) = q\Big((1+q^{-2})T^0 + 0\cdot(qT)\Big) = q+q^{-1}
\ee

If instead of black we would take a white vertex, the orientation of the
hypercube (segment) quiver would be reversed,
\be
C_0 = V, \ \ C_1 = V\otimes V,\nn \\
J(q=1) = -\Big(2 - 2\cdot 2\Big) = 2
\ee
(the minus sign is because every white vertex enters with the coefficient $-q^{-2}$
instead of $q$ for each black vertex).
The differential
\be
d_1 = Q = \left(\begin{array}{cc}
0&0\\  1&0\\  1&0\\  0&1
\end{array}\right)
\ee
has the same rank $2$ and
\be
P(T,q)|_{q=1} = -\Big((2-2)T^0 + (4-2)T^1\Big) = -2T
\ee
Its kernel is empty, but coimage consists of two vectors
$(1000)$ and $(01,-10)$ with $q$-gradings $q^2$ and $1$.
Therefore
\be
P^\circ(T,q) = T^{-1}q^{-2}\Big(0\cdot T^0 + (q^2+1)(qT)\Big) = (q+q^{-1})
\ee

Thus the two superpolynomials $P^\bullet$ and $P^\circ$ coincide
and are equal to the unreduced Jones superpolynomial for the unknot:
\be
P^\bullet(T,q) = P^\circ(T,q) = P^{\rm unknot}(T,q) = q+q^{-1}
\ee

\subsection*{Example: Hopf link}

This is the case of two crossings with two vertices of the same  color.
Then
\be
C_0 = V\otimes V, \ \ \ C_1 = V\oplus V,\ \ \ C_2=V\otimes V, \nn \\
J(q=1) = 2\cdot 2 - (2+2) + 2\cdot 2 = 4
\ee
Here
\be
d_1 =
\left(\begin{array}{c}Q^*\\ \\  \hline\\  Q^*\end{array}\right) =
\left(\begin{array}{cccc}
1 & 0 & 0 & 0 \\ 0 & 1 & 1 & 0 \\ \hline
1 & 0 & 0 & 0 \\ 0 & 1 & 1 & 0 \end{array}\right), \ \ \ \
d_2 = \Big(\begin{array}{c|c}
Q\, &\, -Q \end{array}\Big) =
\left(\begin{array}{cc|cc}
0&0&0&0\\  1&0&-1&0\\  1&0&-1&0\\  0&1&0&-1
\end{array}\right)
\ee
and the ranks and coranks of both operators are equal to $2$,
so that
\be
P(T,q)|_{q=1} = (4-2)T^0 +(4-2-2)T^1 + (4-2) T^2 = 2(1+T^2)
\ee
Explicitly the zero-modes of these operators and their $q$-gradings are:
\be
{\rm Ker} (d_1):    &  (01,-10) \ \& \ (0001) &  1\ \& \ q^{-2} \nn \\
{\rm Im}(d_1):  &   (10|10)\ \& \ (01|01)  & q \ \& \ q^{-1}   \nn \\
{\rm Ker}(d_2):  & (10|10)  \ \& \ (01|01) & q \ \&\ q^{-1}  \nn \\
{\rm CoIm}(d_2): & (01,-10) \ \& \ (1000)&  1\ \& \ q^2
\ee
Therefore
\be
P^{\rm Hopf}(T,q) = q^2\Big((1+q^{-2})T^0 + 0\cdot(qT) + (1+q^2)\cdot(qT)^2\Big)
= (1+q^2)(1+q^4T^2)
\label{PHo}
\ee

\bigskip

If two white vertices were taken instead of the two black ones,
the only difference would be the change of overall factor from $q^2$ to $(Tq^{2})^{-2}$,
i.e. instead of (\ref{PHo}) we obtain
\be
T^{-2}q^{-6}(1+q^2)(1+q^4T^2) = P^{\rm Hopf}(q^{-1},T^{-1})
\ee

\bigskip

If we now consider a pair of vertices of different colors, then
\be
C_0 = V, \ \ \ C_1 = (V\otimes V) \oplus  (V\otimes V), \ \ \ C_2 = V, \nn\\
J(q=1) = \Big(2 - 2\times 2^2 + 2\Big) = 4, \nn \\
\ee
Here
\be
d_1 = \left(\begin{array}{c} Q \\ \hline Q \end{array}\right)
= \left(\begin{array}{cc}
0&0 \\  1&0 \\  1&0 \\  0&1 \\ \hline 0&0 \\  1&0 \\  1&0 \\  0&1
\end{array}\right),
\ee
and
\be
d_2 = \Big(Q^* \Big| -Q^*\Big) =
\left(\begin{array}{cccc|cccc}
1 & 0 & 0 & 0 &   -1 & 0 & 0 & 0 \\
0 & 1 & 1 & 0  &  0 & -1 & -1 & 0 \end{array}\right)
\ee
Both these matrices have rank $2$, therefore
\be
P(T,q)|_{q=1} = T^{-1}\Big((2-2)T^0 + (8-2-2)T^1 + (2-2)T^2\Big) = 4
\ee
Since ${\rm Ker}(d_1) = {\rm CoIm}(d_2) = \emptyset$, the only non-trivial
is the cohomology $H_1$.
The image of $d_1$ consists of two vectors $(0110|0110)$ and $(0001|0001)$,
while the kernel of $d_2$ -- of six: $(abcd|aefg)$, provided $b+c=e+f$.
This leaves in the cohomology $H_1$ just four vectors:
$(1000|0000)$, $(0100|0100)$, $(0100|0010)$, $(0001|0000)$
with the $q$-gradings $q^2$,$1$,$1$,$q^{-2}$ respectively.
Therefore
\be
P(T,q) = \frac{q}{Tq^2}(q^2+2+q^{-2})(qT) = (q+q^{-1})^2
= \Big(P^{\rm unknot}(T,q)\Big)^2
\ee

\subsection*{Example: Trefoil}

We represent trefoil as a 2-srand braid with $3$ black vertices.
Then
\be
C_0 = V\otimes V, \ \ \ C_1 = V\oplus V\oplus V,\ \ \
C_2=(V\otimes V)\oplus (V\otimes V)\oplus (V\otimes V), \ \ \
C_3 = V\otimes V\otimes V \nn \\
J(q=1) = 2\cdot 2 - (2+2+2) + 3(2\cdot 2)-2\cdot 2\cdot 2 = 2, \nn \\
P(q=1,T) = \boxed{2} + (\underline{2}-\underline{2})\cdot T
+ (\underline{\underline{5}}-\underline{\underline{4}})\cdot T^2 + \boxed{1} \cdot T^3
= 2 + T^2 + T^3 \nn \\
P(q,T) = q+q^3 + q^5T^2 + q^9T^3
\ee
This time
\be
d_1 =
\left(\begin{array}{c}Q^* \\   \hline  Q^* \\   \hline  Q^* \end{array}\right),\ \ \ \
d_2 =
\left(\begin{array}{c|c|c}0 & Q & -Q\\ && \\  \hline\hline && \\
Q & 0 & -Q \\ && \\ \hline\hline &&  \\
Q & -Q & 0 \end{array}\right),
\ee
and
\be
d_3 = \left(\begin{array}{c||cc|cc||cc|cc||cc|cc}
&&& \!\!\!\!\!\! V_{23}^{\otimes 2} &&&  & \!\!\!\!\!\!  V_{13}^{\otimes 2}
&&&  & \!\!\!\!\!\! V_{12}^{\otimes 2} \\
&&&&&&&&&&\\
&++&+-&-+&--&++&+-&-+&--&++&+-&-+&--\\
\hline
+++&0&0& && 0&0&& &0&0&& \\
++-&-1&0&&&1&0&&&&&0&0\\
+-+&-1&0&&&&&0&0&-1&0&&\\
+--&0&-1&&&&&1&0&&&-1&0\\
\hline
-++&&&0&0&1&0&&&-1&0&&\\
-+-&&&-1&0&0&1&&&&&-1&0\\
--+&&&-1&0&&&1&0&0&-1&&\\
---&&&0&-1&&&0&1&&&0&-1
\end{array}\right)
\ee
so that $d_3d_2=d_2d_1   = 0$.

For tensor product
\be
{\rm rank}(A\otimes B) = {\rm rank}(A)\cdot{\rm rank}(B)
\ee
Therefore the ranks and coranks of the product matrices are:
\be
\begin{array}{cccc}
{\rm product} & {\rm size} & {\rm rank} & {\rm corank} \\
\hline
d_1\tilde d_1 &  6\times 6 & \underline{2} & 4\\
\tilde d_1 d_1 & 4\times 4 & 2 & \boxed{2}\\
\hline
d_2\tilde d_2 &12\times 12 &\underline{\underline{4}}& 8\\
\tilde d_2d_2 &6\times 6&4&\underline{2}\\
\hline
d_3\tilde d_3 &8\times 8&7&\boxed{1}\\
\tilde d_3d_3 &12\times 12&7&\underline{\underline{5}}
\end{array}
\ee

\bigskip

It remains to restore the $q$-dependence.

The zero-vectors of $\tilde d_1 d_1$ are $v_-\otimes v_-$ and $v_+\otimes v_- - v_-\otimes v_+$,
which have weights $(q^{-1})^2 = q^{-2}$ and $q^{-1}\cdot q = 1$ respectively.

The coimage of $d_3$  is $v_+\otimes v_+\otimes v_+$
and has the weight $q^3\times q^3$ where the second $q^3$ is because it is in the
third term in the complex.

Cohomology $H_1$ is empty. Indeed, $\ {\rm Ker}(d_2)\ $ is spanned by two vectors:
$(v_+,v_+,v_+)$ and $(v_-,v_-,v_-)$ (which would have weights
$q\cdot q=q^2$ and $q^{-1}\cdot q = 1$ respectively)
and they both belong to the image of $d_1$.

Finally, the one-dimensional cohomology $H_2= {\rm Ker}(d_3)/{\rm Im}(d_2)$
is made from the vector
\be
v_+^1\otimes v_-^{23} - v_-^1\otimes v_+^{23} + 2v_+^2\otimes v_-^{13}
+ v_+^3\otimes v_-^{12} - v_-^3 \otimes v_+^{12}
\ee
which has weight $q^{0}\cdot q^2$.

Thus the  superpolynomial is
\be
P(q,T) = q^3\Big( (q^{-2} + 1) + q^2 T^2 + q^6 T^3\Big) =
q+q^3 + q^5T^2 + q^9T^3
\ee

\section*{Acknowledgements}

Our work is partly supported by the
Ministry of Education and Science of the Russian Federation under the contract
2012-1.1-12-000-1003-032 (12.740.11.0677),
by the grants
NSh-3349.2012.2,
RFBR-10-01-00538
and by the joint grants
90453-Ukr, 12-02-91000-ANF, 11-01-92612-Royal Society, 12-02-92108-Yaf-a.

\end{document}